\documentclass[prc,aps,twocolumn,showpacs,amssymb,superscriptaddress]{revtex4}

\usepackage{amsmath}
\usepackage{graphicx}
\usepackage{dcolumn}
\usepackage{float}
\begin{document}

\title{Calculation of Gamow-Teller and Two-Neutrino Double-$\beta$
  Decay Properties for $^{130}$Te and $^{136}$Xe with a realistic
  nucleon-nucleon potential}

\author{L. Coraggio}
\affiliation{Istituto Nazionale di Fisica Nucleare, \\
Complesso Universitario di Monte  S. Angelo, Via Cintia - I-80126 Napoli, Italy}
\author{L. De Angelis}
\affiliation{Istituto Nazionale di Fisica Nucleare, \\
Complesso Universitario di Monte  S. Angelo, Via Cintia - I-80126 Napoli, Italy}
\author{T. Fukui}
\affiliation{Istituto Nazionale di Fisica Nucleare, \\
Complesso Universitario di Monte  S. Angelo, Via Cintia - I-80126 Napoli, Italy}
\author{A. Gargano}
\affiliation{Istituto Nazionale di Fisica Nucleare, \\
Complesso Universitario di Monte  S. Angelo, Via Cintia - I-80126 Napoli, Italy}
\author{N. Itaco}
\affiliation{Istituto Nazionale di Fisica Nucleare, \\ 
Complesso Universitario di Monte  S. Angelo, Via Cintia - I-80126 Napoli, Italy}
\affiliation{Dipartimento di Matematica e Fisica, Universit\`a degli
  Studi della Campania ``Luigi Vanvitelli'', viale Abramo Lincoln 5 -
  I-81100 Caserta, Italy}

\begin{abstract}
We report on the calculation of Gamow-Teller and double-$\beta$ decay
properties for nuclei around $^{132}$Sn within the framework of the
realistic shell model.
The effective shell-model Hamiltonian and Gamow-Teller transition
operator are derived by way of many-body perturbation theory, without
resorting to empirical effective quenching factor for the Gamow-Teller
operator.
The results are then compared with the available experimental data, in
order to establish the reliability of our approach.
This is a mandatory step, before we apply the same methodology, in
forthcoming studies, to the calculation of the neutrinoless
double-$\beta$ decay nuclear matrix element for nuclei that are
currently considered among the best candidates for the detection of
this process.
\end{abstract}


\maketitle

\section{Introduction}
\label{intro}
The detection of neutrinoless double-$\beta$ decay ($0\nu\beta\beta$)
is nowadays one of the main targets in many laboratories all around
the world, triggered by the search of ``new physics'' beyond the
Standard Model.
The observation of such a process would be the evidence of a lepton
number violation and shed more light on the nature and properties of
the neutrino (see Refs. \cite{Avignone08,Vergados12} and references
therein).

It is well known that the expression for the half life of the
$0\nu\beta\beta$ decay can be written in the following form:

\begin{equation}
\left[ T^{0\nu}_{1/2} \right]^{-1} = G^{0\nu} \left| M^{0\nu}
\right|^2 \langle m _{\nu} \rangle^2 ~~,
\label{halflife}
\end{equation}

\noindent
where $G^{0\nu}$ is the so-called phase-space factor (or kinematic
factor), $\langle m _{\nu} \rangle$ is the effective neutrino
mass that takes into account the neutrino parameters associated
with the mechanisms of light- and heavy-neutrino exchange, and
$M^{0\nu}$ is the nuclear matrix element (NME) directly related to the
wave functions of the parent and grand-daughter nuclei.

From the expression (\ref{halflife}), it is clear that a reliable
estimate of the NME is a keypoint to understand which are the
most favorable nuclides to be considered for the search of the
$0\nu\beta\beta$ decay, and how to link the experimental results to a
measurement of $|\langle m_{\nu}\rangle|$.
It is therefore incumbent upon the theoretical nuclear structure
community to make an effort to provide calculations of the NME as much
reliable as possible.

Currently, the nuclear structure models which are largely employed in
this research field are the Interacting Boson Model (IBM)
\cite{Barea09,Barea12,Barea13}, the Quasiparticle Random-Phase
Approximation (QRPA) \cite{Simkovic08,Simkovic09,Fang11,Faessler12},
Energy Density Functional methods \cite{Rodriguez10}, and the Shell
Model (SM)
\cite{Caurier08,Menendez09a,Menendez09b,Horoi13a,Horoi13b,Neacsu15,Brown15,Frekers17}. 
All of them have different advantages and drawbacks, that make one
model more suitable than another for a certain class of nuclei, but
nowadays the results obtained employing these approaches agree within a
factor $\sim 2 \div 3$ (see Ref. \cite{Barea15} and references therein).

A common feature in all the many-body models applied to systems with
mass number ranging from $A=48$ to 150 is that the parameters upon
which they depend need to be determined fitting some spectroscopic
properties of the nuclei under investigation.
In particular, since the Hilbert space considered in these
approximated models is a truncated one, it is necessary to introduce
quenching factors of the axial and vector coupling constants $g_A$ and
$g_V$ that appear in the NME expression.
Besides of the excluded degrees of freedom in the many-body
calculation, the quenching operation has to take into account the
subnucleonic structure of the nucleons too.
The free value of  $g_A$, that is obtained by the measurement of
$g_A/g_V$ from the neutron decay \cite{Nakamura10}, is 1.269, and its
quenching factor is usually fixed fitting the observed Gamow-Teller
(GT) and two-neutrino double-$\beta$ decay ($2\nu\beta\beta$)
properties, that are experimentally available.

We remark that the structure of the two operators, corresponding to
the $0\nu\beta\beta$ and $2\nu\beta\beta$ decays,  is quite different,
and the quenching operation may be effective to calculate the GT
strengths and $2\nu\beta\beta$ NME, but not consistent with the
renormalization of the $0\nu\beta\beta$-decay operator.

As a matter of fact, there are two main open questions about this
problem.
The first one is related to the fact that in the $2\nu\beta\beta$
decay essentially the $J^{\pi}=1^+$ states of the intermediate odd-odd
nucleus are involved in the process, while all multipoles come into
play in the $0\nu\beta\beta$ decay.
So there is no precise prescription if $0\nu\beta\beta$ should be
quenched only for the $1^+$ multipole, the quenching factor being
fitted on $\beta$-decay properties, or all the multipole channels
should be equally quenched \cite{Barea13}.

Besides this, there is another question to be addressed.
In the $2\nu\beta\beta$ decay the term associated with the vector
current of the electroweak lagrangian and its coupling constant $g_V$
plays a negligible role, but this might not be the case for the
$0\nu\beta\beta$ decay.
So, it may be necessary also to renormalize this factor in order to
take into account the many-body effects and the neglected subnucleonic
degrees of freedom.
Actually, there is no experimental evidence for an underlying
mechanism for the renormalization of $g_V$, namely if the same
quenching factor used for $g_A$, fixed by fitting $\beta$ decay data,
should be used to quench $g_V$ too.

Our framework to tackle these problems is the realistic shell model,
where all the parameters appearing in the SM Hamiltonian and in the
transition operators are derived from a realistic free nucleon-nucleon
($NN$) potential $V_{NN}$ by way of the many-body theory
\cite{Kuo90,Suzuki95}.
In this way the bare matrix elements of the $NN$ potential and of any
transition operator are renormalized with respect to the truncation of
the full Hilbert space into the reduced SM model space, to take
into account the neglected degrees of freedom without resorting to
any empirical parameter.
In other words, in our approach we do not employ effective charges to
calculate electromagnetic transition strengths, and we do not quench
empirically the axial and vector current coupling constants.

It is a mandatory step, however, to check this approach to calculate
properties related to the GT and $2\nu\beta\beta$ decays of
nuclei involved in possible $0\nu\beta\beta$, and compare the results
with the available data.
This is the content of present work, where we present the outcome of
SM calculations for nuclei around $^{132}$Sn, focussing our attention
on the GT strengths and $2\nu\beta\beta$-decay of $^{130}$Te and
$^{136}$Xe.

These two nuclei are currently considered as candidates for the
observation of neutrinoless double-beta decay by some large
experimental collaborations.
The $0\nu\beta\beta$ decay of $^{130}$Te is targeted by the CUORE
collaboration at the INFN Laboratori Nazionali del Gran Sasso in Italy
\cite{CUORE}, while the decay of $^{136}$Xe is investigated both by
the EXO-200 collaboration at the Waste Isolation Pilot Plant in
Carlsbad, New Mexico, \cite{EXO-200}, and by the KamLAND-Zen
collaboration in the Kamioka mine in Japan \cite{Kamland}.

Our starting point is the high-precision $NN$ potential CD-Bonn
\cite{Machleidt01b}, whose repulsive high-momentum components are
smoothed out using the $V_{\rm  low-k}$ approach \cite{Bogner02}.
Then, from this realistic potential we have derived, within a model
space spanned by the five proton and neutron orbitals
$0g_{7/2},1d_{5/2},1d_{3/2},2s_{1/2},0h_{11/2}$ outside the
doubly-closed $^{100}$Sn, the effective shell-model Hamiltonian $H_{\rm
  eff}$, effective electromagnetic and GT transition operators.
The derivation of the effective Hamiltonian and operators has been
performed by way of the time-dependent perturbation theory
\cite{Kuo71,Coraggio09a}, including diagrams up to the third-order in
$V_{\rm low-k}$.

The following section is devoted to the presentation of some details
about the derivation of our shell-model Hamiltonian and of the effective
transition and decay operators.
In Section \ref{results}, we report the results of our calculations
for the spectroscopic properties of $^{130}$Te, $^{130,136}$Xe, and
$^{136}$Ba, electromagnetic and GT transition strengths for
$^{130}$Te and $^{136}$Xe, and their NMEs for the $2\nu\beta\beta$
decay.
Theoretical results are compared with available experimental data.
In the last section we sketch out a summary of the present work and an
outlook of our future program.
In the Supplemental Material \cite{supplemental2017}, the calculated
two-body matrix elements (TBME) of our SM Hamiltonian can be found.

\section{Outline of calculations}
\label{calculations}
We start our calculations by considering the high-precision CD-Bonn
$NN$ potential \cite{Machleidt01b}.
Because of the non-perturbative behavior induced by the repulsive
high-momentum components of CD-Bonn potential, we have renormalized
the latter by way of the so-called $V_{\rm low-k}$ approach
\cite{Bogner01,Bogner02}.
This procedure provides a smooth potential that can be employed
directly in the many-body perturbation theory, and that preserves
exactly the onshell properties of the original $NN$ potential up to a
cutoff momentum $\Lambda$.
We have chosen its value, as in many of our recent papers
\cite{Coraggio09c,Coraggio09d,Coraggio14b,Coraggio15a,Coraggio16a}, to
be equal to $2.6$ fm$^{-1}$, because we have found that the larger the
cutoff the smaller the role of the missing three-nucleon force (3NF)
\cite{Coraggio15b}.
The Coulomb potential is explicitly taken into account in the
proton-proton channel.

The next step is to derive an effective Hamiltonian for SM
calculations employing a model space spanned by the five 
$0g_{7/2},1d_{5/2},1d_{3/2},2s_{1/2},0h_{11/2}$
proton and neutron orbitals outside the doubly-closed $^{100}$Sn
core.
To this end, an auxiliary one-body potential $U$ is introduced in
order to break up the Hamiltonian for a system of $A$ nucleons as the
sum of a one-body term $H_0$, which describes the independent motion
of the nucleons, and a residual interaction $H_1$:

\begin{eqnarray}
 H &= & \sum_{i=1}^{A} \frac{p_i^2}{2m} + \sum_{i<j=1}^{A} V_{\rm low-k}^{ij}
 = T + V_{\rm low-k} = \nonumber \\
~& = & (T+U)+(V_{\rm low-k}-U)= H_{0}+H_{1}~~.\label{smham}
\end{eqnarray}

\noindent
Once $H_0$ has been introduced, the reduced model space is defined in
terms of a finite subset of $H_0$'s eigenvectors.
In our calculation we choose as auxiliary potential the harmonic
oscillator (HO) potential.

Since the diagonalization of the many-body Hamiltonian (\ref{smham}) in an
infinite Hilbert space is obviously infeasible, our eigenvalue problem is then
reduced to the solution of that one for an effective 
Hamiltonian $H_{\rm eff}$ in a truncated model space.

In this paper, we derive $H_{\rm eff}$ by way of the Kuo-Lee-Ratcliff
(KLR) folded-diagram expansion \cite{Kuo71,Kuo90} in terms of the
vertex function $\hat{Q}$ box, that is defined as

\begin{equation}
\hat{Q} (\epsilon) = P H_1 P + P H_1 Q \frac{1}{\epsilon - Q H Q} Q
H_1 P ~~. \label{qbox}
\end{equation}

\noindent
The $\hat{Q}$ box may be expanded perturbatively in terms of
irreducible valence-linked one- and two-body Goldstone diagrams
through third order in $H_1$ \cite{Hjorth95}.
We have reviewed the calculation of our SM effective Hamiltonian
$H_{\rm eff}$ in Ref. \cite{Coraggio12a}, where details of the
diagrammatic expansion of the $\hat{Q}$ box and its perturbative
properties are also reported.

In terms of the $\hat{Q}$ box, the effective SM Hamiltonian $H_{\rm
  eff}$ can be written in an operator form as 

\begin{equation}
H_{\rm eff} = \hat{Q} - \hat{Q'} \int \hat{Q} + \hat{Q'} \int \hat{Q} \int
\hat{Q} - \hat{Q'} \int \hat{Q} \int \hat{Q} \int \hat{Q} + ~...~~,
\end{equation}

\noindent
where the integral sign represents a generalized folding operation, 
and $\hat{Q'}$ is obtained from $\hat{Q}$ by removing terms at the
first order in $V_{\rm low-k}$ \cite{Kuo71,Kuo90}.
The folded-diagram series is then summed up to all orders using the
Lee-Suzuki iteration method \cite{Suzuki80}.

From $H_{\rm eff}$ we obtain both single-particle (SP) energies and
TBME for our SM calculations.
As already mentioned in the Introduction, in the Supplemental Material
\cite{supplemental2017} our calculated TBME are reported, and in Table
\ref{spetab} our calculated SP energies.
There, the latter (labelled as I) are compared with a set of empirical
SP energies (labelled as II) that are needed to fit the observed SP
states in $^{133}$Sb and $^{131}$Sn \cite{ensdf,xundl}.

\begin{table}[ht]
\caption{Theoretical (I) and empirical (II) proton and neutron SP energy
  spacings (in MeV) employed in present work (see text for details).}
\begin{ruledtabular}
\begin{tabular}{cccccc}
\label{spetab}
 ~ & ~ & Proton SP spacings & ~ & ~ &  Neutron SP spacings \\
   ~ & I & II & ~ & I & II \\
\colrule
 $0g_{7/2}$   & 0.0 & 0.0 &~& 0.0 & 0.0 \\ 
 $1d_{5/2}$   & 0.3 & 0.4 &~& 0.6 & 0.7 \\ 
 $1d_{3/2}$   & 1.2 &1.4 &~& 1.5 & 2.1 \\ 
 $2s_{1/2}$   & 1.1 &1.3 &~& 1.2 & 1.9 \\ 
 $0h_{11/2}$ & 1.9 & 1.6 &~& 2.7 & 3.0 \\ 
\end{tabular}
\end{ruledtabular}
\end{table}

As regards the effective transition and decay operators, namely the
effective charges of the electric quadrupole operators and the matrix
elements of the effective GT operator, we have derived them
consistently with SM $H_{\rm eff}$, within an approach that is
strictly based on the one presented by Suzuki and Okamoto in
Ref. \cite{Suzuki95}.

In that paper, it has been demonstrated that a non-Hermitian effective
operator $\Theta_{\rm eff}$ can be written in the following form:

\begin{eqnarray}
\Theta_{\rm eff} & = & (P + \hat{Q}_1 + \hat{Q}_1 \hat{Q}_1 + \hat{Q}_2
\hat{Q} + \hat{Q} \hat{Q}_2 + \cdots)(\chi_0+\nonumber \\
~ & ~& + \chi_1 + \chi_2 +\cdots)~~, \label{effopexp}
\end{eqnarray}

\noindent
where $\hat{Q}$ is the $\hat{Q}$ box defined by the expression (\ref{qbox}), and 

\begin{equation}
\hat{Q}_m = \frac {1}{m!} \frac {d^m \hat{Q} (\epsilon)}{d \epsilon^m} \biggl| 
_{\epsilon=\epsilon_0} ~~, 
\label{qm}
\end{equation}

\noindent
$\epsilon_0$ being the eigenvalue of the degenerate model-space of
the unperturbed Hamiltonian $H_0$, that, as mentioned before, we have
chosen to be the HO one.

The $\chi_n$ operators are defined as follows:

\begin{eqnarray}
\chi_0 &=& (\hat{\Theta}_0 + h.c.)+ \Theta_{00}~~,  \label{chi0} \\
\chi_1 &=& (\hat{\Theta}_1\hat{Q} + h.c.) + (\hat{\Theta}_{01}\hat{Q}
+ h.c.) ~~, \\
\chi_2 &=& (\hat{\Theta}_1\hat{Q}_1 \hat{Q}+ h.c.) +
(\hat{\Theta}_{2}\hat{Q}\hat{Q} + h.c.) + \nonumber \\
~ & ~ & (\hat{\Theta}_{02}\hat{Q}\hat{Q} + h.c.)+  \hat{Q}
\hat{\Theta}_{11} \hat{Q}~~, \label{chin} \\
&~~~& \cdots \nonumber
\end{eqnarray}

\noindent
where $\hat{\Theta}_m$, $\hat{\Theta}_{mn}$ have the following
expressions:
\begin{eqnarray}
\hat{\Theta}_m & = & \frac {1}{m!} \frac {d^m \hat{\Theta}
 (\epsilon)}{d \epsilon^m} \biggl|_{\epsilon=\epsilon_0} ~~~, \\
\hat{\Theta}_{mn} & = & \frac {1}{m! n!} \frac{d^m}{d \epsilon_1^m}
\frac{d^n}{d \epsilon_2^n}  \hat{\Theta} (\epsilon_1 ;\epsilon_2)
\biggl|_{\epsilon_1= \epsilon_0, \epsilon_2  = \epsilon_0} ~,
\end{eqnarray}

\noindent
with
\begin{eqnarray}
\hat{\Theta} (\epsilon) & = & P \Theta P + P \Theta Q
\frac{1}{\epsilon - Q H Q} Q H_1 P ~~, \label{thetabox} \\
\hat{\Theta} (\epsilon_1 ; \epsilon_2) & = & P \Theta P + P H_1 Q
\frac{1}{\epsilon_1 - Q H Q} \times \nonumber \\
~ & ~ & Q \Theta Q \frac{1}{\epsilon_2 - Q H Q} Q H_1 P ~~,
\end{eqnarray}

\noindent
$\Theta$ being the bare operator.

In our calculations for the one-body operators we arrest the $\chi$
series to the leading term $\chi_0$, and the latter is expanded
perturbatively including diagrams up to the third order in the
perturbation theory, consistently with the perturbative expansion of
the $\hat{Q}$ box.

\begin{figure}[ht]
\begin{center}
\includegraphics[scale=0.40,angle=0]{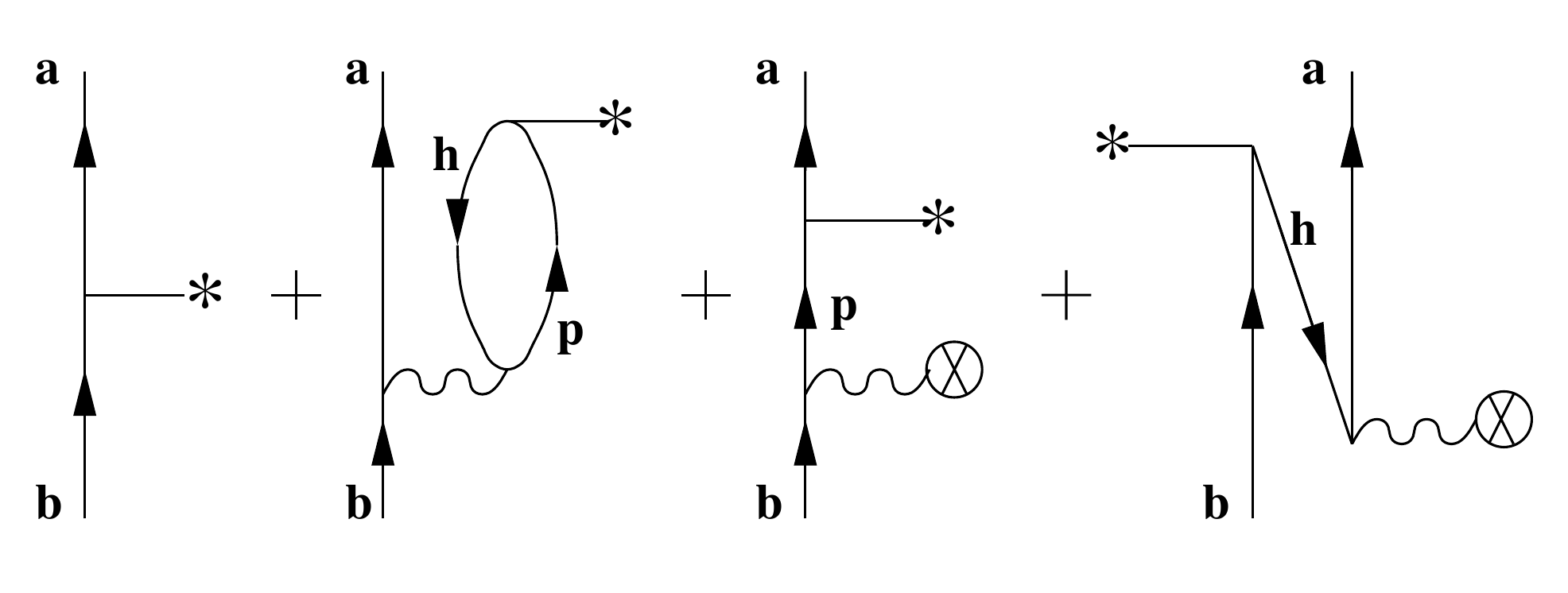}
\caption{One-body second-order diagrams included in the perturbative
  expansion of $\chi_0$. The asterisk indicates the bare operator
  $\Theta$, the wavy lines the two-body potential $V_{\rm low-k}$.}
\label{figeffop}
\end{center}
\end{figure}

In Fig. \ref{figeffop} we have reported all the single-body $\chi_0$
diagrams up to the second order, the bare operator $\Theta$ being
represented with an asterisk.
The first-order $(V_{\rm low-k}-U)$-insertion is represented by a
circle with a cross inside, which arises in the perturbative expansion
owing to the presence of the $−U$ term in the interaction Hamiltonian
$H_1$ (see for example Ref. \cite{Coraggio12a} for details).

Using this approach we have calculated proton and neutron effective
state-dependent charges, which are reported in Table \ref{effch}.
It should be pointed out that our results are close to the usual empirical
values ($e^{\rm emp}_p=1.5e,~e^{\rm emp}_n =0.5\div 0.8e$).

\begin{table}[ht]
\caption{Proton and neutron effective charges of the electric
  quadrupole operator $E2$.}
\begin{ruledtabular}
\begin{tabular}{ccc}
\label{effch}
$n_a l_a j_a ~ n_b l_b j_b $ &  $\langle a || e_p || b \rangle $ &
$\langle a || e_n || b \rangle $ \\
\colrule
 $0g_{7/2}~0g_{7/2}$     & 1.66 & 1.00 \\ 
 $0g_{7/2}~1d_{5/2}$     & 1.70 & 1.07 \\ 
 $0g_{7/2}~1d_{3/2}$     & 1.65 & 1.00 \\ 
 $1d_{5/2}~0g_{7/2}$     & 1.71 & 1.00 \\ 
 $1d_{5/2}~1d_{5/2}$     & 1.52 & 0.63 \\ 
 $1d_{5/2}~1d_{3/2}$     & 1.50 & 0.64 \\ 
 $1d_{5/2}~2s_{1/2}$     & 1.53 & 0.62 \\ 
 $1d_{3/2}~0g_{7/2}$     & 1.63 & 0.97 \\ 
 $1d_{3/2}~1d_{5/2}$     & 1.48 & 0.66 \\ 
 $1d_{3/2}~1d_{3/2}$     & 1.51 & 0.69 \\ 
 $1d_{3/2}~2s_{1/2}$     & 1.55 & 0.68 \\ 
 $2s_{1/2}~1d_{5/2}$     & 1.52 & 0.63 \\ 
 $2s_{1/2}~1d_{3/2}$     & 1.56 & 0.67 \\ 
 $0h_{11/2}~0h_{11/2}$  & 1.50 & 0.68 \\ 
\end{tabular}
\end{ruledtabular}
\end{table}

In Tables \ref{effGTpn} and \ref{effGTnp}, the matrix elements of the
proton-neutron GT$^+$ and neutron-proton GT$^-$ effective operators,
respectively, are reported.

The breaking of the proton-neutron symmetry is due to the fact that we
include in the perturbative calculation of $H_{\rm eff}$ and ${\rm
  GT}_{\rm eff}$ also the effect of the Coulomb potential between the
interacting protons.
In the last column the quenching factors that should be employed in
order to obtain the corresponding ${\rm GT}_{\rm eff}$ matrix element
are also reported.
The quenching factor is not reported for those matrix elements that
are forbidden for the bare GT operator.

\begin{table}[ht]
\caption{Matrix elements of the proton-neutron effective GT$^+$
  operator. In the last column it is reported the corresponding
  quenching factors (see text for details).}
\begin{ruledtabular}
\begin{tabular}{ccc}
\label{effGTpn}
$n_a l_a j_a ~ n_b l_b j_b $ & GT$^+_{\rm eff}$ &  quenching
factor \\
\colrule
 $0g_{7/2}~0g_{7/2}$     & -1.239 & 0.50 \\ 
 $0g_{7/2}~1d_{5/2}$     & -0.139 & ~ \\ 
 $1d_{5/2}~0g_{7/2}$     & 0.017 & ~ \\ 
 $1d_{5/2}~1d_{5/2}$     & 1.864 & 0.64 \\ 
 $1d_{5/2}~1d_{3/2}$     & -1.747 & 0.56 \\ 
 $1d_{3/2}~1d_{5/2}$     & 1.942 & 0.63 \\ 
 $1d_{3/2}~1d_{3/2}$     & -1.023 & 0.66 \\ 
 $1d_{3/2}~2s_{1/2}$     & -0.118 & ~ \\ 
 $2s_{1/2}~1d_{3/2}$     & 0.095 & ~ \\ 
 $2s_{1/2}~2s_{1/2}$     & 1.598 & 0.65 \\ 
 $0h_{11/2}~0h_{11/2}$  & 2.597 & 0.69 \\ 
\end{tabular}
\end{ruledtabular}
\end{table}

\begin{table}[ht]
\caption{Same as in Table \ref{effGTpn}, but for the neutron-proton
  effective GT$^-$ operator.}
\begin{ruledtabular}
\begin{tabular}{ccc}
\label{effGTnp}
$n_a l_a j_a ~ n_b l_b j_b $ &  GT$^-_{\rm eff}$ &  quenching
factor \\
\colrule
 $0g_{7/2}~0g_{7/2}$     & -1.239 & 0.50 \\ 
 $0g_{7/2}~1d_{5/2}$     & -0.019 & ~ \\ 
 $1d_{5/2}~0g_{7/2}$     & 0.131 & ~ \\ 
 $1d_{5/2}~1d_{5/2}$     & 1.864 & 0.64 \\ 
 $1d_{5/2}~1d_{3/2}$     & -1.891 & 0.61 \\ 
 $1d_{3/2}~1d_{5/2}$     & 1.794 & 0.58 \\ 
 $1d_{3/2}~1d_{3/2}$     & -1.023 & 0.66 \\ 
 $1d_{3/2}~2s_{1/2}$     & -0.093 & ~ \\ 
 $2s_{1/2}~1d_{3/2}$     & 0.117 & ~ \\ 
 $2s_{1/2}~2s_{1/2}$     & 1.598 & 0.65 \\ 
 $0h_{11/2}~0h_{11/2}$  & 2.597 & 0.69 \\ 
\end{tabular}
\end{ruledtabular}
\end{table}

\section{Results}\label{results}

This section is devoted to the presentation of the results of our SM
calculations.

We compare the calculated low-energy spectra of $^{130}$Te,
$^{130}$Xe, $^{136}$Xe, and $^{136}$Ba, and their electromagnetic
transition strengths with the available experimental data, that are
reported in Table \ref{E2}.
It should be mentioned that in Ref. \cite{Vietze15} shell-model
calculations for $^{130,136}$Xe isotopes have been performed using the
empirical shell-model Hamiltonian GCN5082 \cite{Menendez09b}.

We show also the results of the GT$^-$ strength distributions of
$^{130}$Te and $^{136}$Xe, which are defined as follows:

\begin{equation}
B({\rm GT}^-) = \frac{ \left| \langle \Phi_f || \sum_{j}
  \vec{\sigma}_j \tau^-_j || \Phi_i \rangle \right|^2} {2J_i+1}~~,
\label{GTstrength}
\end{equation}

\noindent
where indices $i,f$ refer to the parent and daughter nuclei,
respectively, and the sum is over all interacting nucleons.

In the following subsections, we also report the results of the
calculated NME of the $2\nu\beta\beta$ decays $^{130}{\rm Te}_{\rm
  g.s.} \rightarrow ^{130}$Xe$_{\rm g.s.}$ and $^{136}{\rm Xe}_{\rm
  g.s.} \rightarrow ^{136}$Ba$_{\rm g.s.}$, via the following expression:

\begin{equation}
M^{\rm GT}_{2\nu} = \sum_n \frac{ \langle 0^+_f || \vec{\sigma} \tau^-
  || 1^+_n \rangle \langle 1^+_n || \vec{\sigma}
\tau^- || 0^+_i \rangle } {E_n + E_0} ~~,
\label{doublebetame}
\end{equation}

\noindent
where $E_n$ is the excitation energy of the $J^{\pi}=1^+_n$
intermediate state, $E_0=\frac{1}{2}Q_{\beta\beta}(0^+) +\Delta M$,
$Q_{\beta\beta}(0^+)$ and $\Delta M$ being the $Q$ value of the $\beta
\beta$ decay and the mass difference between the daughter and parent
nuclei, respectively.
In the expression of Eq. (\ref{doublebetame}) the sum over index $n$
runs over all possible intermediate states of the daughter nucleus.
The NMEs have been calculated using the ANTOINE shell-model code,
using the Lanczos strength function-method as in Ref. \cite{Caurier05}.
The theoretical values are then compared with the experimental
counterparts, that are directly related to the observed
half life $T^{2\nu}_{1/2}$ 

\begin{equation}
\left[ T^{2\nu}_{1/2} \right]^{-1} = G^{2\nu} \left| M^{\rm GT}_{2\nu}
\right|^2 ~~.
\label{2nihalflife}
\end{equation}

In connection with the $2\nu\beta\beta$ decay, we show also the
comparison between our calculated proton/neutron occupancies/vacancies
and the recent data.

All the calculations have been performed employing both theoretical
and empirical SP energies, reported in Table \ref{spetab}, in order to
provide an indicator of the sensitivity of our SM results on the
choice of the SP energies.

\begin{table}[ht]
\caption{Experimental and calculated $B(E2)$ strengths of $^{130}$Te,
  $^{130}$Xe, $^{136}$Xe, and $^{136}$Ba (in $e^2{\rm fm}^4$). They are
  reported for observed states up to 2 MeV excitation energy. Data are
  taken from Refs \cite{ensdf,xundl}.}
\begin{ruledtabular}
\begin{tabular}{ccccc}
\label{E2}
 Nucleus & $J_i \rightarrow J_f $ & $B(E2)_{Expt}$ & I & II \\
\colrule
 ~ & ~ & ~ & ~ & ~ \\
 $^{130}$Te   &          ~                     & ~ & ~ & ~ \\
~                 & $2^+ \rightarrow 0^+$ & $580 \pm 20$ & 430 & 420  \\
~                 & $6^+ \rightarrow 4^+$ & $240 \pm 10$ & 220 & 200  \\
 $^{130}$Xe   &          ~                     & ~ & ~ & ~ \\
~                 & $2^+ \rightarrow 0^+$ & $1170^{+20}_{-10}$ & 954 & 876  \\
 $^{136}$Xe   &          ~                     & ~ & ~ & ~ \\
~                 & $2^+ \rightarrow 0^+$ & $420 \pm 20$ & 300 & 300  \\
~                 & $4^+ \rightarrow 2^+$ & $53 \pm 1$ & 9 & 11  \\
~                 & $6^+ \rightarrow 4^+$ & $0.55 \pm 0.02$ & 1.58 & 2.42  \\
 $^{136}$Ba   &          ~                     & ~ & ~ & ~ \\
~                 & $2^+ \rightarrow 0^+$ & $800^{+80}_{-40}$ & 590 & 520
\end{tabular}
\end{ruledtabular}
\end{table}

\subsection{$^{130}$Te ${\rm GT}^-$ strengths and $2\nu\beta\beta$
  decay}

\begin{center}
\begin{figure}[ht]
\includegraphics[scale=0.44,angle=0]{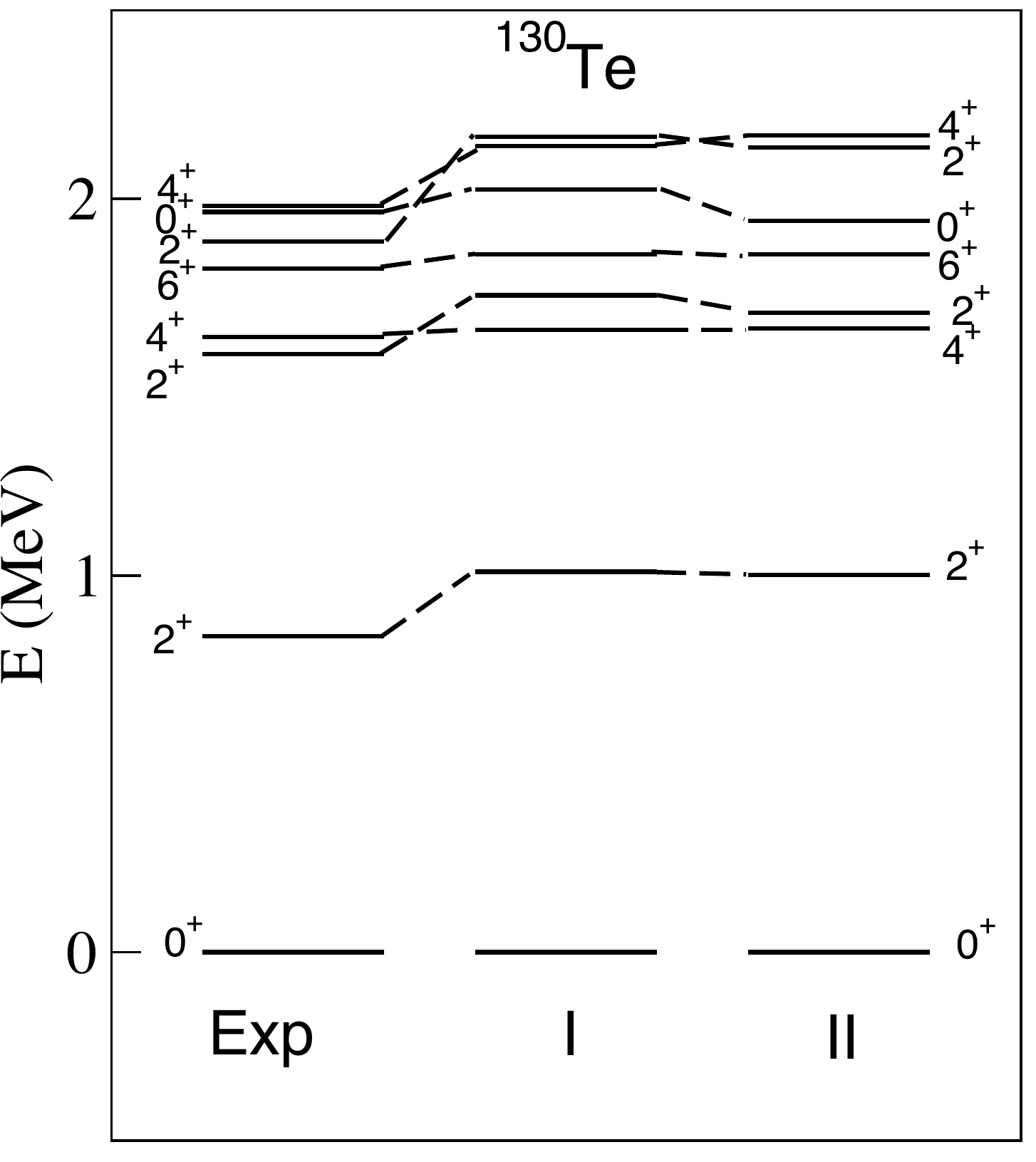}
\caption{Experimental and calculated spectra of $^{130}$Te up to 2 MeV
  excitation energy (see text for details).}
\label{130Te}
\end{figure}
\end{center}

\begin{center}
\begin{figure}[ht]
\includegraphics[scale=0.44,angle=0]{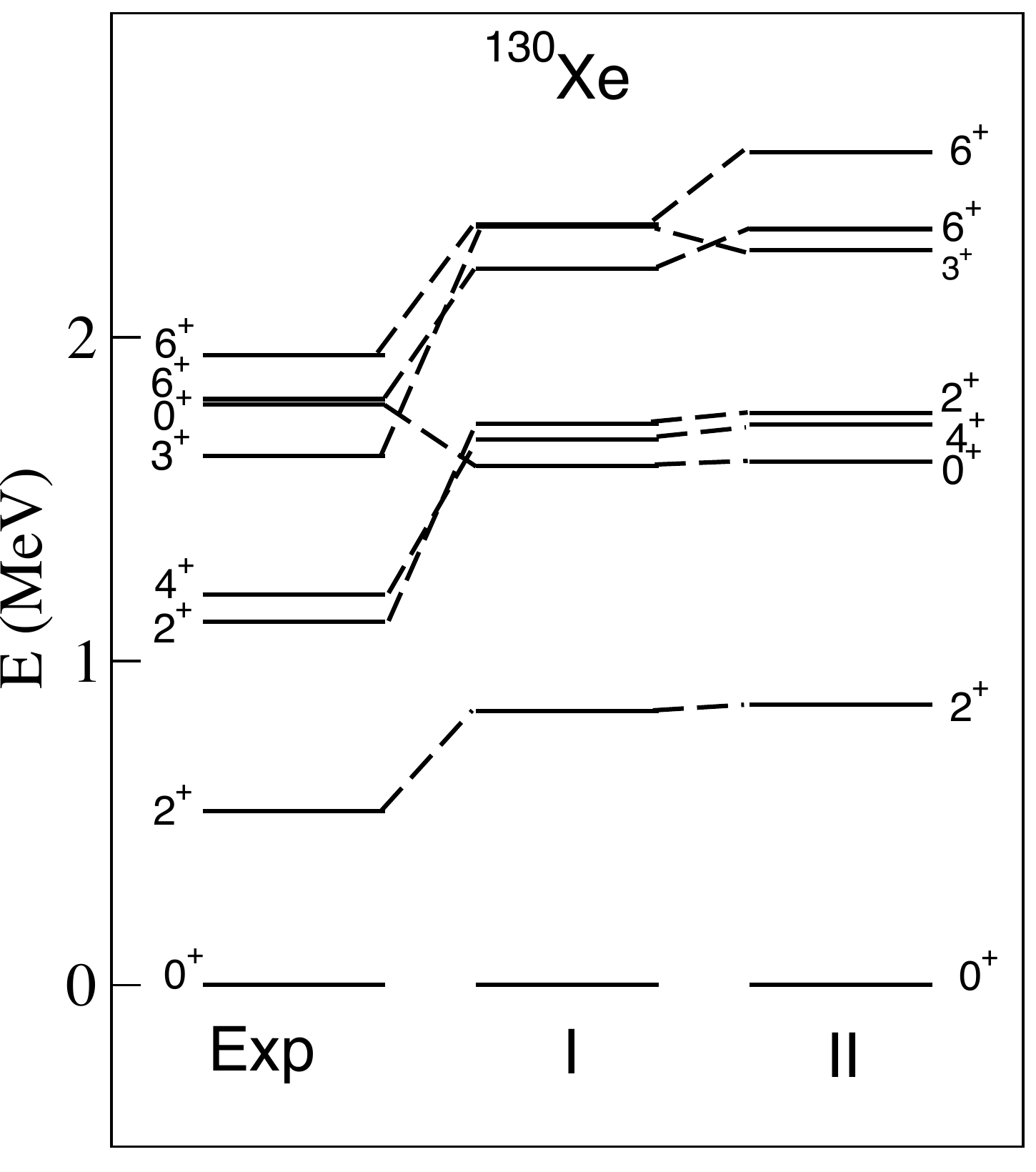}
\caption{Same as Fig. \ref{130Te}, for $^{130}$Xe.}
\label{130Xe}
\end{figure}
\end{center}

In Figs. \ref{130Te} and \ref{130Xe}, we show the experimental
\cite{ensdf,xundl} and calculated spectra of $^{130}$Te and $^{130}$Xe
up to an excitation energy of 2 MeV.
As can be seen, these results are scarcely sensitive to the choice of
the SP energies, those of $^{130}$Te being in a very good agreement
with the experimental data, while the reproduction of the observed
$^{130}$Xe low-lying states is less satisfactory.

From inspection of Table \ref{E2}, it can be seen that our calculated
electric-quadrupole transition rates $B(E2)$ compare well with the
observed values for both nuclei, testifying the reliability of our SM
wavefunctions and of the effective electric-quadrupole
transition-operator. 
Its matrix elements are reported in Table \ref{effch}.
It is worth noting that the calculated $B(E2)$s do not show a
relevant dependence on the choice of the SP energies, their values
being very close each other.

\begin{center}
\begin{figure}[ht]
\includegraphics[scale=0.08,angle=0]{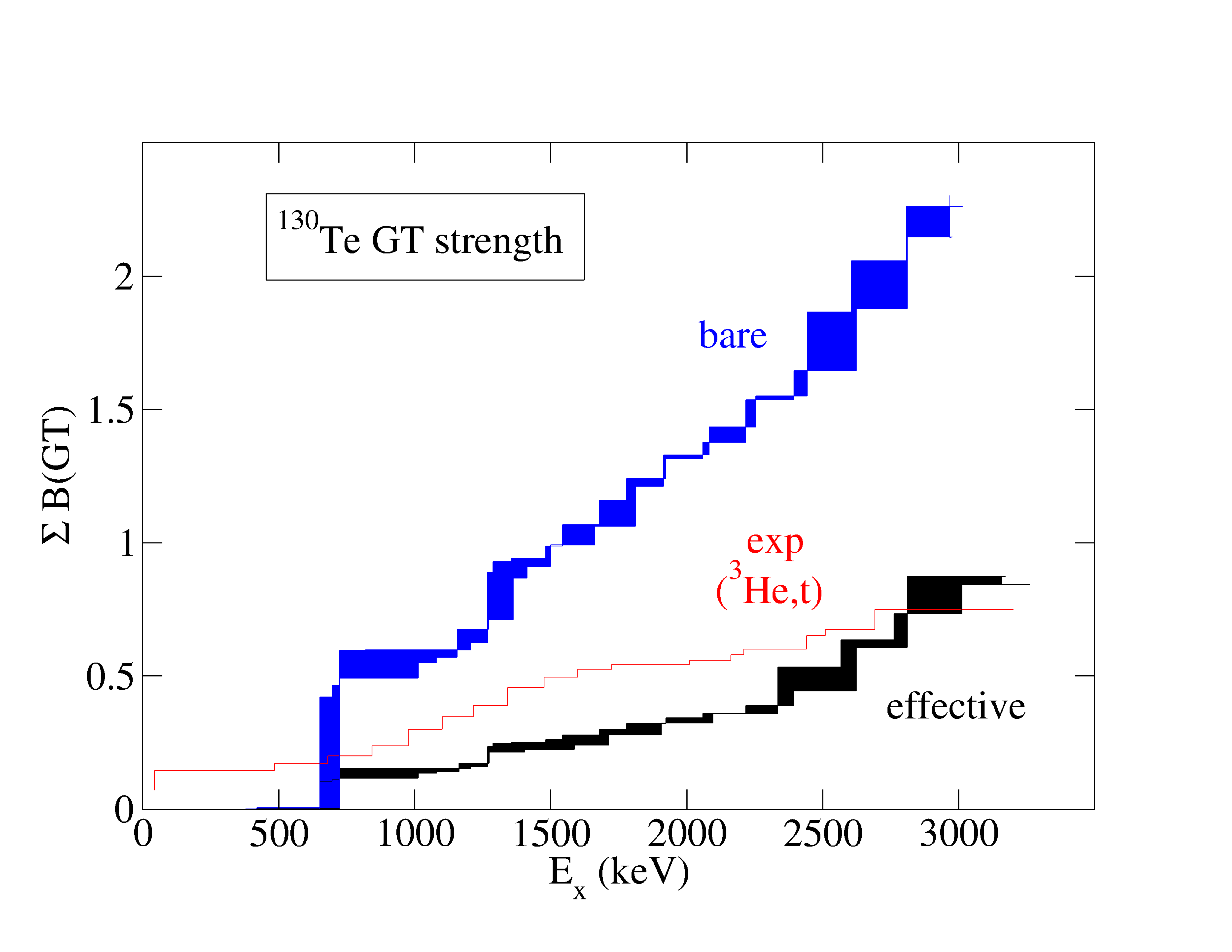}
\caption{Running sums of the $^{130}$Te $B({\rm GT}^-)$
  strengths as a function of the excitation energy $E_x$ up to 3000
  keV (see text for details).}
\label{130TeGT-}
\end{figure}
\end{center}

In Fig. \ref{130TeGT-}, our calculated running sums of the
Gamow-Teller strengths ($\Sigma B({\rm GT}^-)$) as a function of the
excitation energy for $^{130}$Te are shown.
The comparison of the calculated GT strength distributions with the
observed ones is a very relevant point when trying to assess the
reliability of a many-body approach to the description of the $\beta\beta$
decay.

The single $\beta$ decay GT strengths, defined by
Eq. (\ref{GTstrength}), can be accessed experimentally through
intermediate energy charge-exchange reactions.
As a matter of fact, the GT strength can be extracted, following the
standard approach in the distorted-wave Born approximation (DWBA),
from the GT component of the cross section by way of the relation
\cite{Goodman80,Taddeucci87}

\begin{equation}
\frac{d \sigma^{GT}}{d \Omega} = \left (\frac{\mu}{\pi \hbar^2} \right
)^2 \frac{k_f}{k_i} N^{\sigma \tau}_{D}| J_{\sigma \tau} |^2 B(GT)~~,
\end{equation}

\noindent
where $N^{\sigma \tau}_{D}$ is the distortion factor, and $| J_{\sigma
  \tau} |$ is the volume integral of the effective $NN$ interaction.

In the following, we compare our results with the GT$^-$ distributions
obtained in recent high-resolution $(^3{\rm He},t)$ studies on
$^{130}$Te \cite{Puppe12}.

In Fig. \ref{130TeGT-}, the data are reported with a red line,
while the results obtained with SP energies (I) and (II) define 
the blue and black areas, respectively, for the bare and
effective ${\rm GT}^-$ operators.
It can be seen that the renormalized GT operator is able to reproduce
quite well the behavior of the experimental running GT strength.

As a matter of fact, if we shift the calculated distributions in order
to reproduce the position of the first $1^+$ state in $^{130}$I,
the theoretical total ${\rm GT}^-$ strengths up to 3 MeV excitation
energy are equal to 0.842 and 0.873,  for the calculations with SP
energies I and II respectively, which should be compared with the
experimental value $0.746 \pm 0.045$.
The crucial role of the many-body renormalization is evident when
considering the results obtained using the bare GT operator.
In this case the total ${\rm GT}^-$ strength is equal to 2.554 and
2.408 with SP energies from set (I) and (II), respectively.

\begin{center}
\begin{figure}[ht]
\includegraphics[scale=0.08,angle=0]{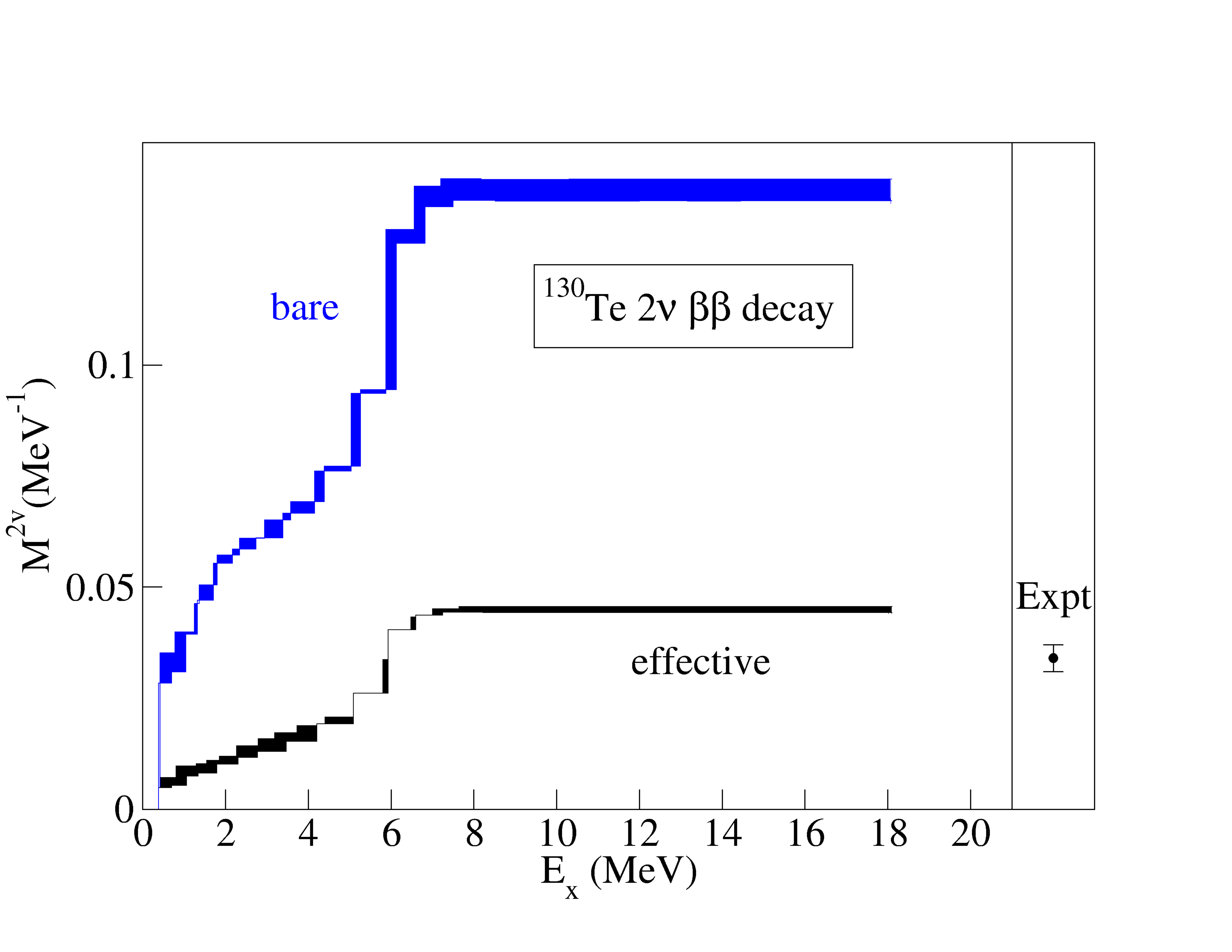}
\caption{Running sums of the calculated $M^{\rm
    GT}_{2\nu}$ as a function of the excitation energy of the
  intermediate states. The blue area corresponds to the calculations
  with the bare GT operator, while the black one to those with ${\rm
    GT}_{\rm eff}$ (see text for details).}
\label{130Te130Xe}
\end{figure}
\end{center}

As regards the $2\nu\beta\beta$ decay of $^{130}$Te,
we have calculated NME, as defined by expression (\ref{doublebetame}),
and the results, compared with value obtained from the experimental
half life of the $^{130}{\rm Te} \rightarrow ^{130}$Xe
$2\nu\beta\beta$ decay \cite{Barabash10}, are reported in
Fig. \ref{130Te130Xe}.

The theoretical results are reported as a function of the maximum
excitation energy of the intermediate states included in the sum of
expression (\ref{doublebetame}).
As can be seen, the calculated values saturate when including at least
intermediate states up to 8 MeV excitation energy.

As in the case of the theoretical GT strength distributions, the NMEs
calculated with the effective GT operator are in a good
agreement with the experimental datum $M^{\rm GT}_{2\nu}=(0.034 \pm
0.003)$MeV$^{-1}$ \cite{Barabash10}, our results being 0.044 MeV$^{-1}$
and 0.046 MeV$^{-1}$ with SP energies (I) and (II), respectively.
Actually, the NMEs calculated with the bare GT
operator are 0.131 MeV$^{-1}$(I) and 0.137 MeV$^{-1}$ (II), which are
far away from the experimental one.

It is worth mentioning now the results obtained by two recent SM
calculations \cite{Caurier12,Neacsu15}, where the shell model
Hamiltonians are based on realistic $NN$ potentials but empirically
modified in order to reproduce some spectroscopic properties of nuclei
around $^{132}$Sn.
In Ref. \cite{Caurier12} the calculated NME is 0.043 MeV$^{-1}$, with
a quenching factor of 0.57 of $g_A$, while in Ref. \cite{Neacsu15} a
value of 0.0328 MeV$^{-1}$, using a quenching factor of 0.74, is
reported.

\begin{center}
\begin{figure}[ht]
\includegraphics[scale=0.40,angle=0]{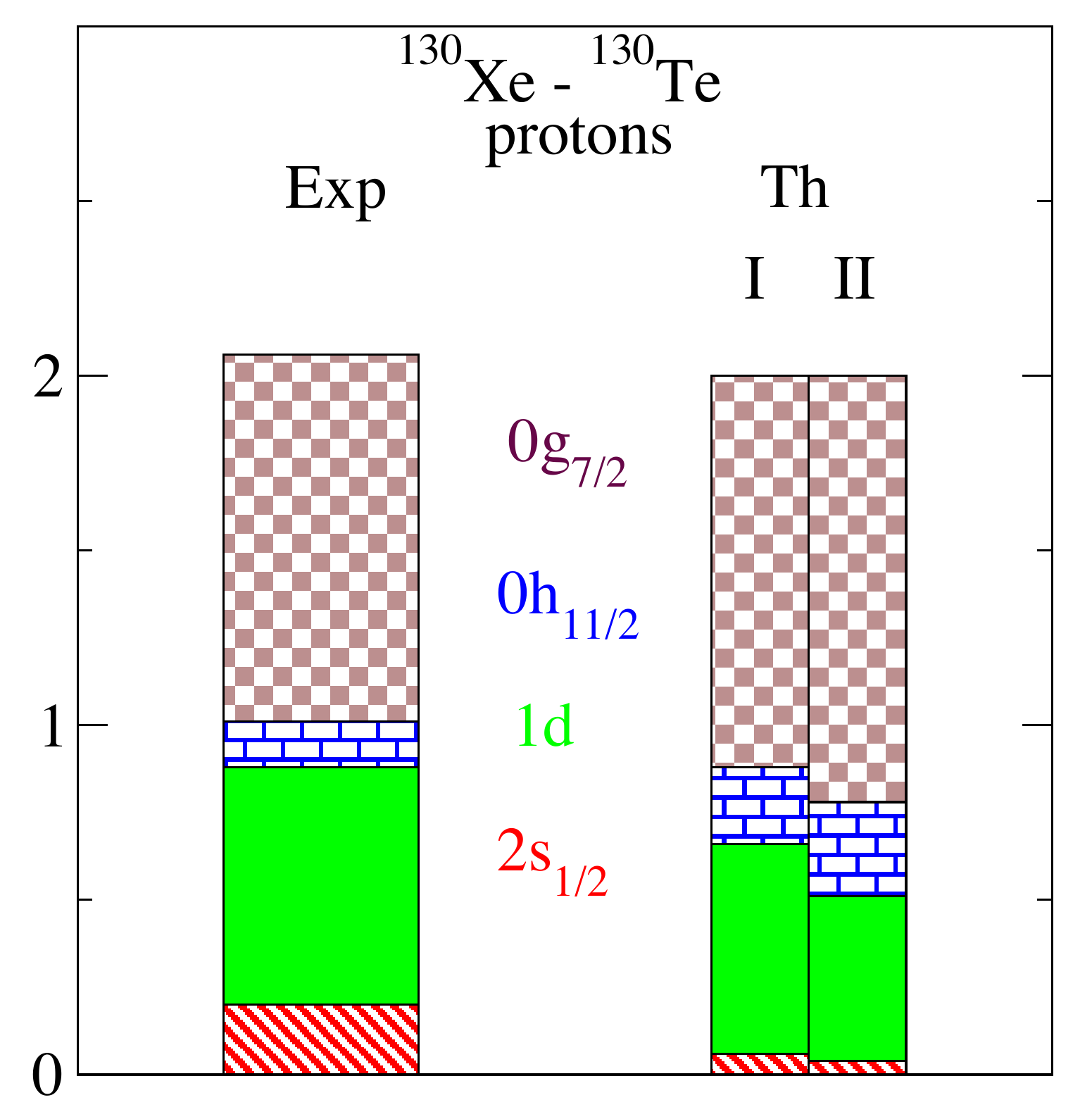}
\caption{Change in proton occupancies between the
  ground states for the $^{130}{\rm Te} \rightarrow ^{130}$Xe
  decay (see text for details). The brown area corresponds to the
  occupation of the $0g_{7/2}$ orbital, the green one to the $1d$
  orbitals, the red one to the $2s_{1/2}$ orbital, and the blue one to
  the $0h_{11/2}$ orbital.}
\label{130Teprot}
\end{figure}
\end{center}

\begin{center}
\begin{figure}[ht]
\includegraphics[scale=0.40,angle=0]{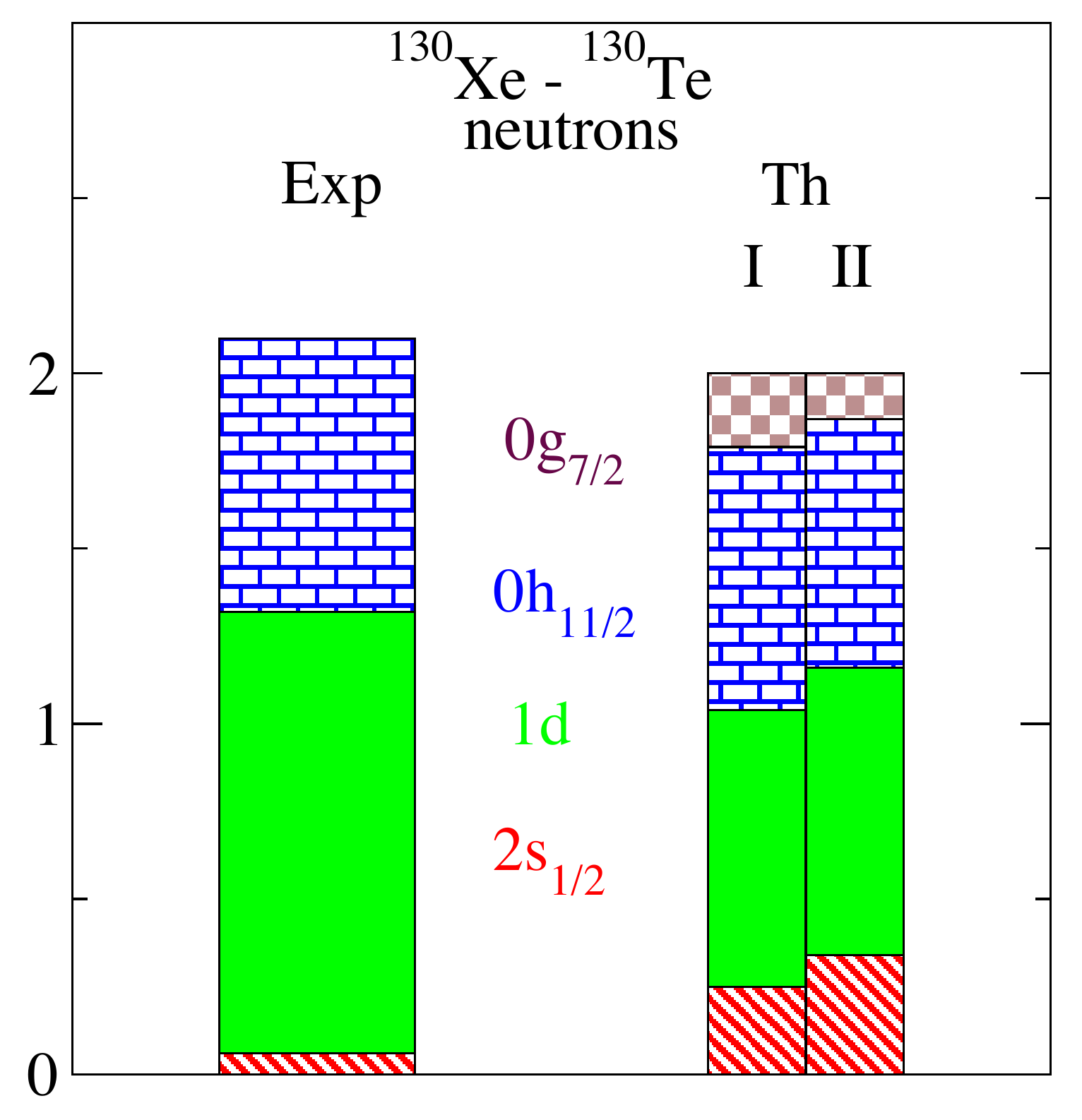}
\caption{Change in neutron vacancies between the
  ground states for the $^{130}{\rm Te} \rightarrow ^{130}$Xe
  decay (see text for details). Colored areas refer to the same
  orbital as in Fig. \ref{130Teprot}.}
\label{130Teneut}
\end{figure}
\end{center}

Another important indicator of the quality of the calculated NME, both
for $2\nu\beta\beta$ and $0\nu\beta\beta$ decay, may be provided by
the comparison of the theoretical occupancies of valence nucleons in
the ground states of the parent and grand-daughter nuclei with the
observed ones.
Recently, those quantities have been determined by measuring the cross
sections of one-proton stripping and one-neutron pick-up reactions, for
proton occupancies and neutron vacancies, respectively
\cite{Entwisle16,Kay13}.
These data are reported in Figs. \ref{130Teprot} and \ref{130Teneut} and
compared with our calculations.

\noindent
The calculations with SP energies (I) and (II) give very close
results, which are in nice agreement with experiment, bearing in mind
the experimental uncertainties that are up to $20\%$ for the change in
occupancy of proton $0g_{7/2}$ orbital \cite{Entwisle16}.

\subsection{$^{136}$Xe ${\rm GT}^-$ strengths and $2\nu\beta\beta$
  decay}

\begin{center}
\begin{figure}[ht]
\includegraphics[scale=0.44,angle=0]{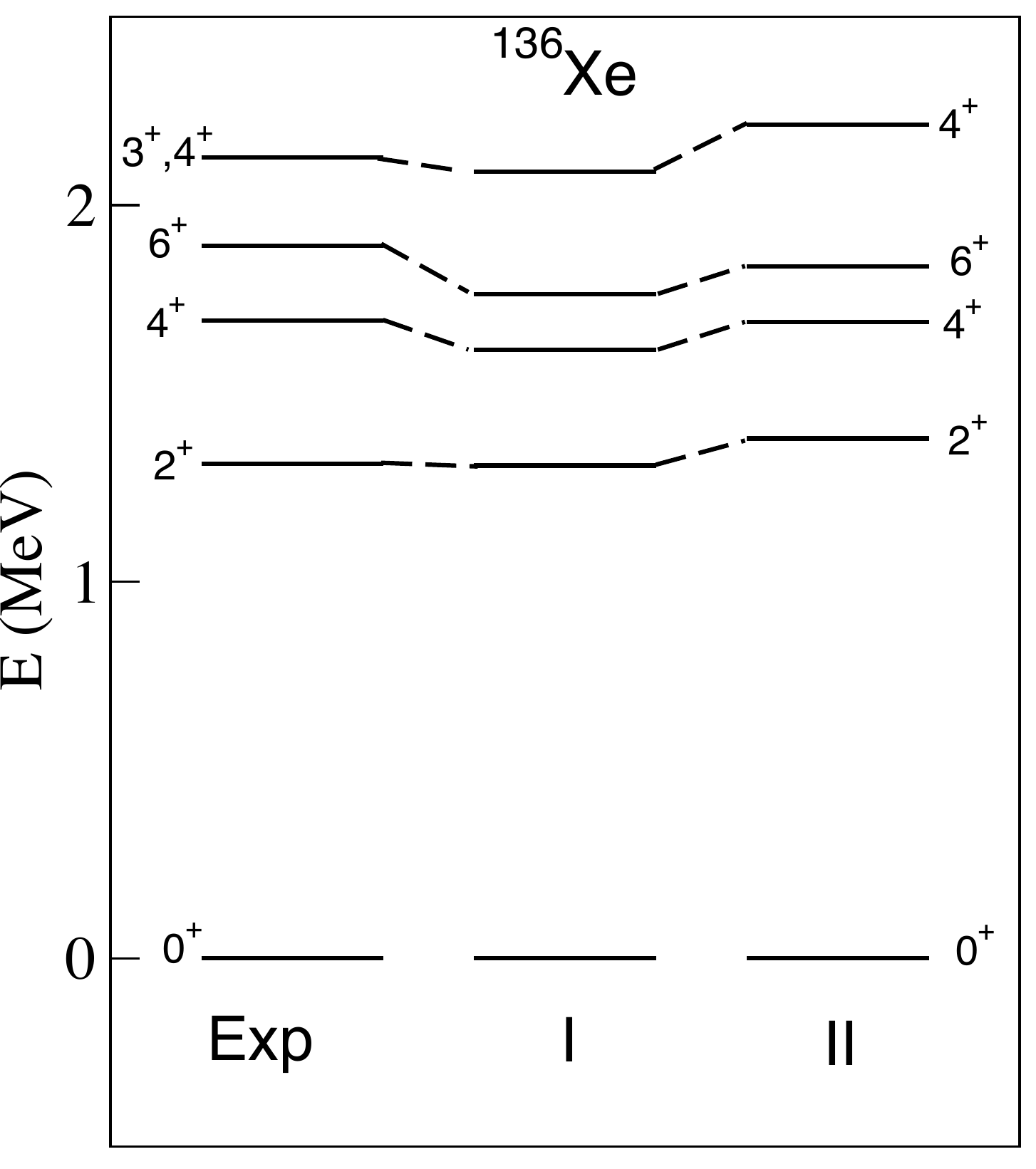}
\caption{Same as Fig. \ref{130Te}, for $^{136}$Xe.}
\label{136Xe}
\end{figure}
\end{center}
 
\begin{center}
\begin{figure}[ht]
\includegraphics[scale=0.44,angle=0]{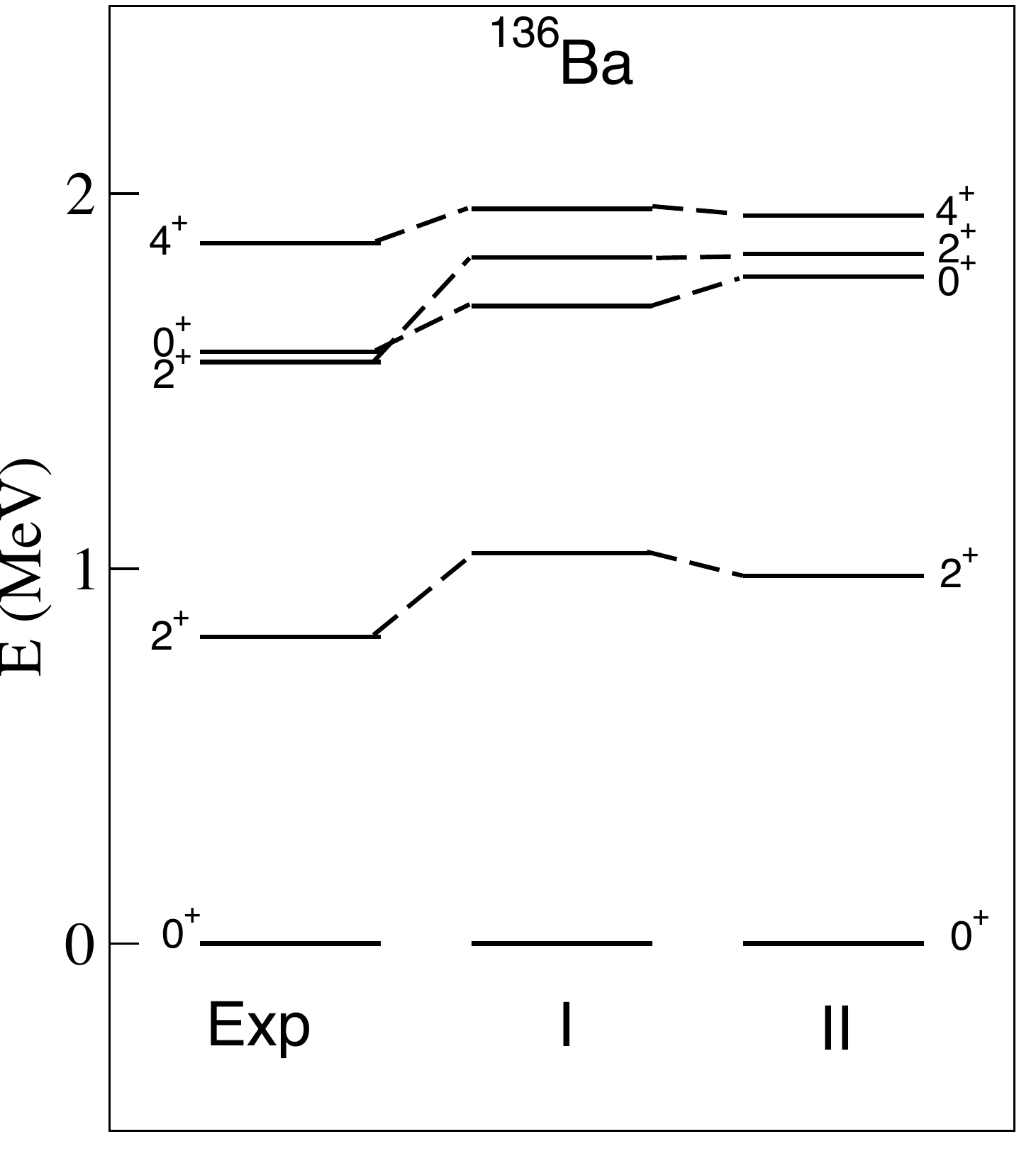}
\caption{Same as Fig. \ref{130Te}, for $^{136}$Ba.}
\label{136Ba}
\end{figure}
\end{center}

This subsection is organized like the previous one, so we start from the
inspection of Figs. \ref{136Xe} and \ref{136Ba}, where the
experimental \cite{ensdf,xundl} and calculated spectra of $^{136}$Xe
and $^{136}$Ba up to an excitation energy of 2 MeV are reported.
The calculated spectra are again in a good agreement with experiment,
and results are rather insensitive to the SP energies I and II.

From inspection of Table \ref{E2}, it can be seen that our calculated
$B(E2;2^+_1 \rightarrow 0^+_1)$s are very close to the observed value,
while for $^{136}$Xe the theoretical $B(E2;4^+_1 \rightarrow 2^+_1)$s
and $B(E2;6^+_1 \rightarrow 4^+_1)$s are less satisfactory when
compared with available data \cite{ensdf,xundl}, calculations with SP
energies (I) and (II) underestimating the observed $B(E2;4^+_1
\rightarrow 2^+_1)$ and overestimating the experimental $B(E2;6^+_1
\rightarrow 4^+_1)$.

\begin{center}
\begin{figure}[ht]
\includegraphics[scale=0.08,angle=0]{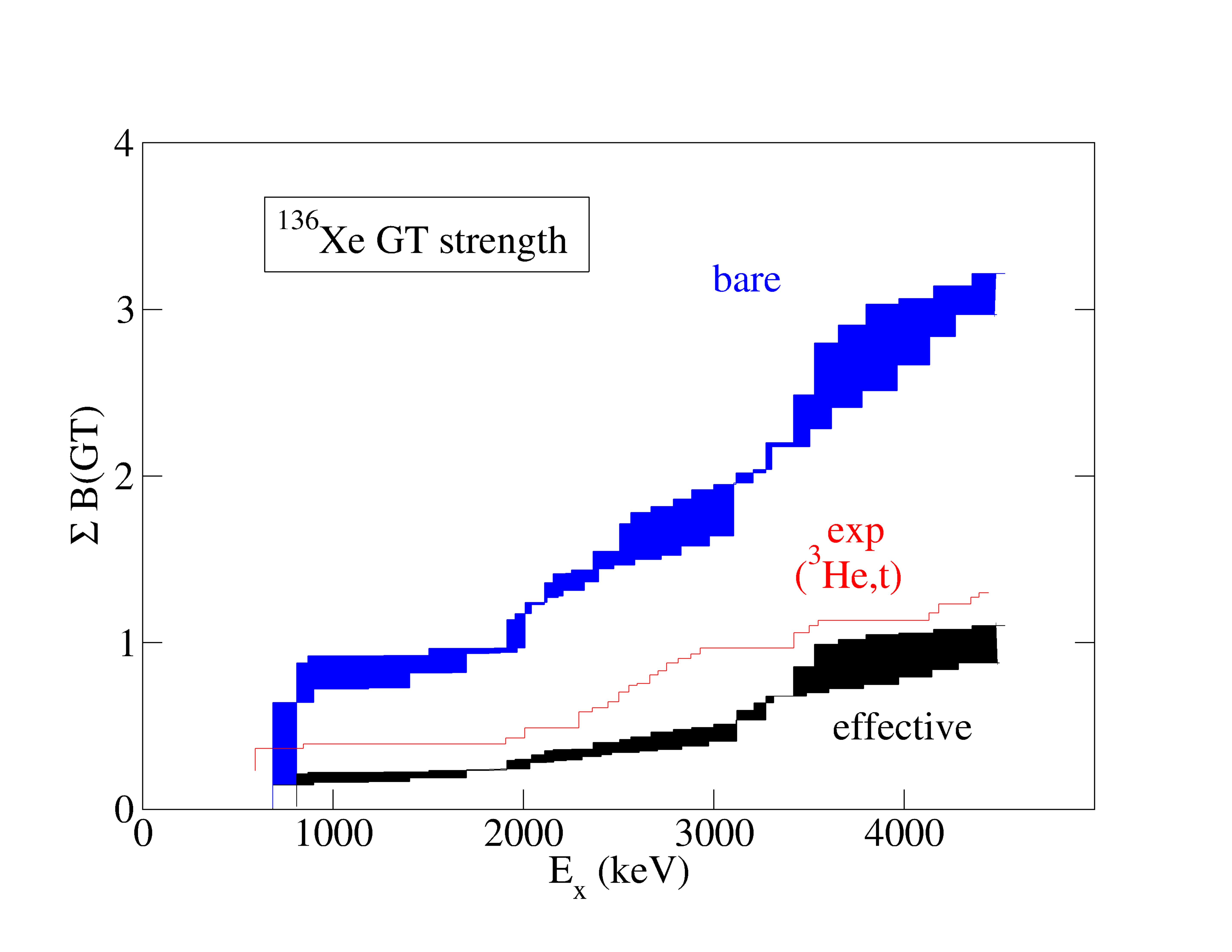}
\caption{Running sums of the $^{136}$Xe $B({\rm GT}^-)$ strengths as a
  function of the excitation energy $E_x$ up to 4500 keV (see text for
  details).}
\label{136XeGT-}
\end{figure}
\end{center}

The calculated $\Sigma B({\rm GT}^-)$ for $^{136}$Xe, as a function of
the excitation energy, can be found in Fig. \ref{136XeGT-}, where they
are compared with the observed GT$^-$ distributions extracted from 
high-resolution $(^3{\rm He},t)$ reactions on $^{136}$Xe \cite{Frekers13}.

As in the case of $^{130}$Te, we observe that the renormalized GT
operator reproduces satisfactorily the observed running GT strength,
the theoretical total ${\rm GT}^-$ strengths up to 4.5 MeV excitation
energy being equal to 0.94 (I) and 1.13 (II) and to be compared with
an experimental value of $1.33 \pm 0.07$.

\begin{center}
\begin{figure}[ht]
\includegraphics[scale=0.08,angle=0]{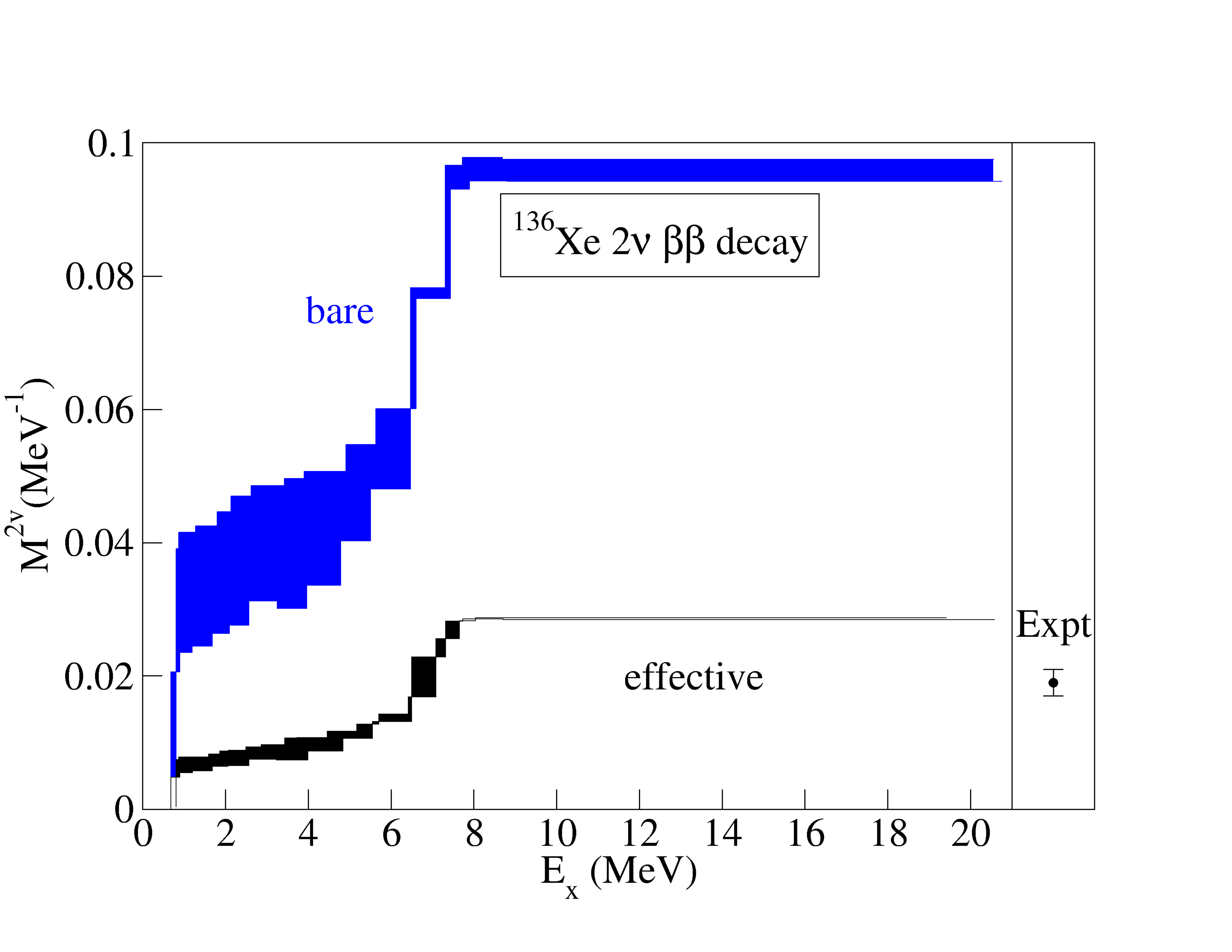}
\caption{Same as in Fig. \ref{130Te130Xe}, for the
  $^{136}{\rm Xe} \rightarrow ^{136}$Ba $2\nu\beta\beta$ decay
  (see text for details).}
\label{136Xe136Ba}
\end{figure}
\end{center}

We have calculated the NME related to the $2\nu\beta\beta$ decay of
$^{136}$Xe into $^{136}$Ba, whose values are 0.091 MeV$^{-1}$ (I) and
0.094 MeV$^{-1}$ (II) with bare GT operator, and 0.0285 MeV$^{-1}$ (I)
and 0.0287 MeV$^{-1}$ (II) with the effective operator $\rm{GT}_{\rm
  eff}$.
The experimental value, obtained from the experimental half life of
the $^{136}{\rm Xe} \rightarrow ^{136}$Ba $2\nu\beta\beta$ decay
\cite{Albert14}, is $(0.0218 \pm 0.0003)$ MeV$^{-1}$, that
compares well with the theoretical values derived employing the
effective operator (Fig. \ref{136Xe136Ba}).

For the sake of completeness, we mention that in Ref. \cite{Caurier12}
the calculated value of the NME is 0.025 MeV$^{-1}$ with a quenching
factor equal to 0.45, and in Ref. \cite{Neacsu15} they obtain 0.0256
MeV$^{-1}$, having quenched $g_A$ by a factor 0.74.
The latter has been employed also in Ref. \cite{Horoi13b} as a quenching
factor to calculate the matrix element of $2\nu\beta\beta$ decay of
$^{136}$Xe, resulting in a calculated NME of 0.062 MeV$^{-1}$.
In this paper, the authors have used a different shell-model
Hamiltonian, derived by way of the KLR folded-diagram expansion from
the realistic N$^3$LO potential \cite{Entem02}, which however seems to
describe the nuclear structure of nuclei around $Z=50$ equally as well
as in Ref. \cite{Neacsu15}.
This evidences the tight relationship between the shell-model Hamiltonian
and the choice of the $g_A$ quenching factor.

\begin{center}
\begin{figure}[ht]
\includegraphics[scale=0.40,angle=0]{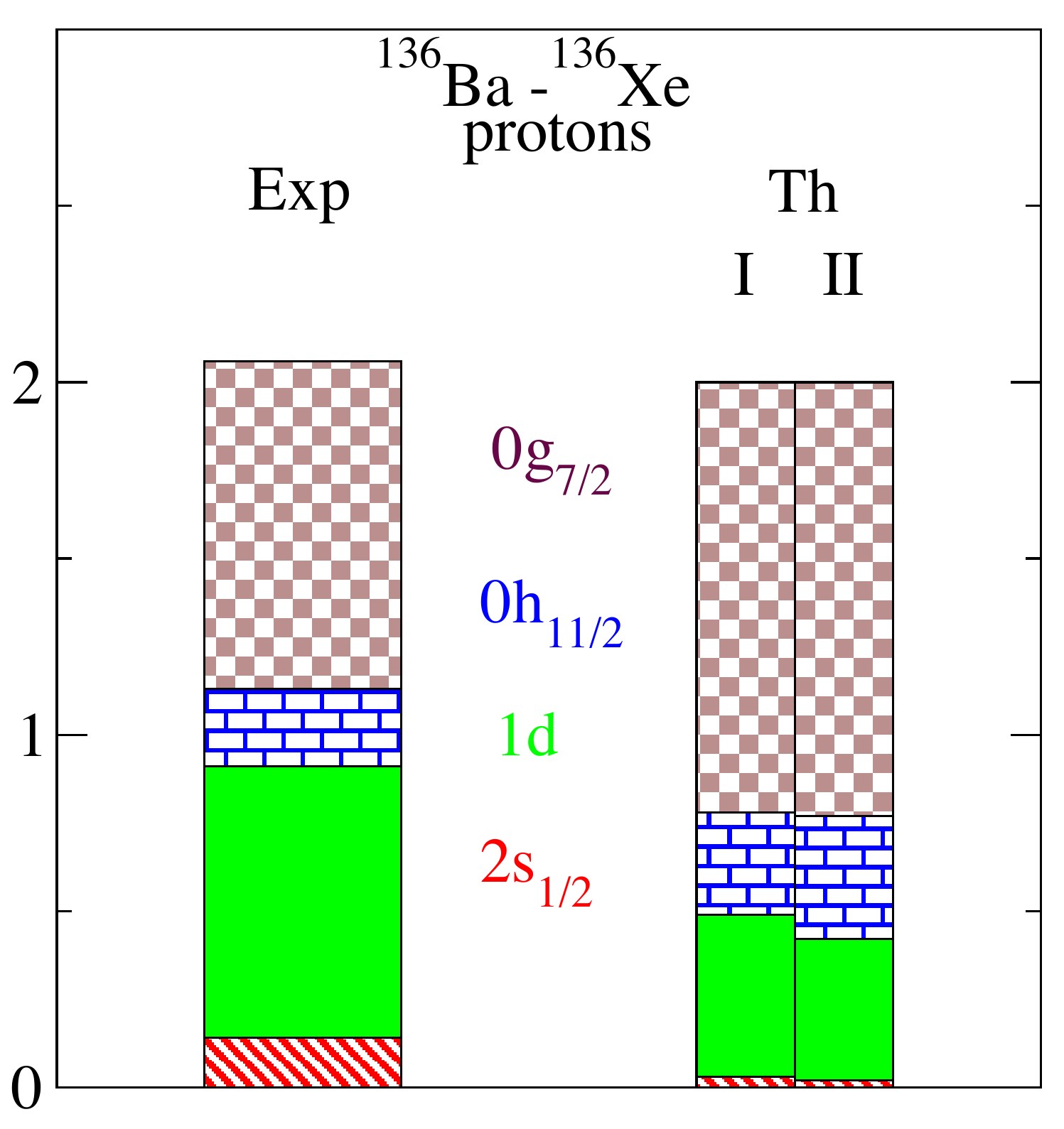}
\caption{Same as in Fig. \ref{130Teprot}, for the
  $^{136}{\rm Xe}\rightarrow ^{136}$Ba decay (see text for
  details).}
\label{136Xeprot}
\end{figure}
\end{center}

\begin{center}
\begin{figure}[ht]
\includegraphics[scale=0.40,angle=0]{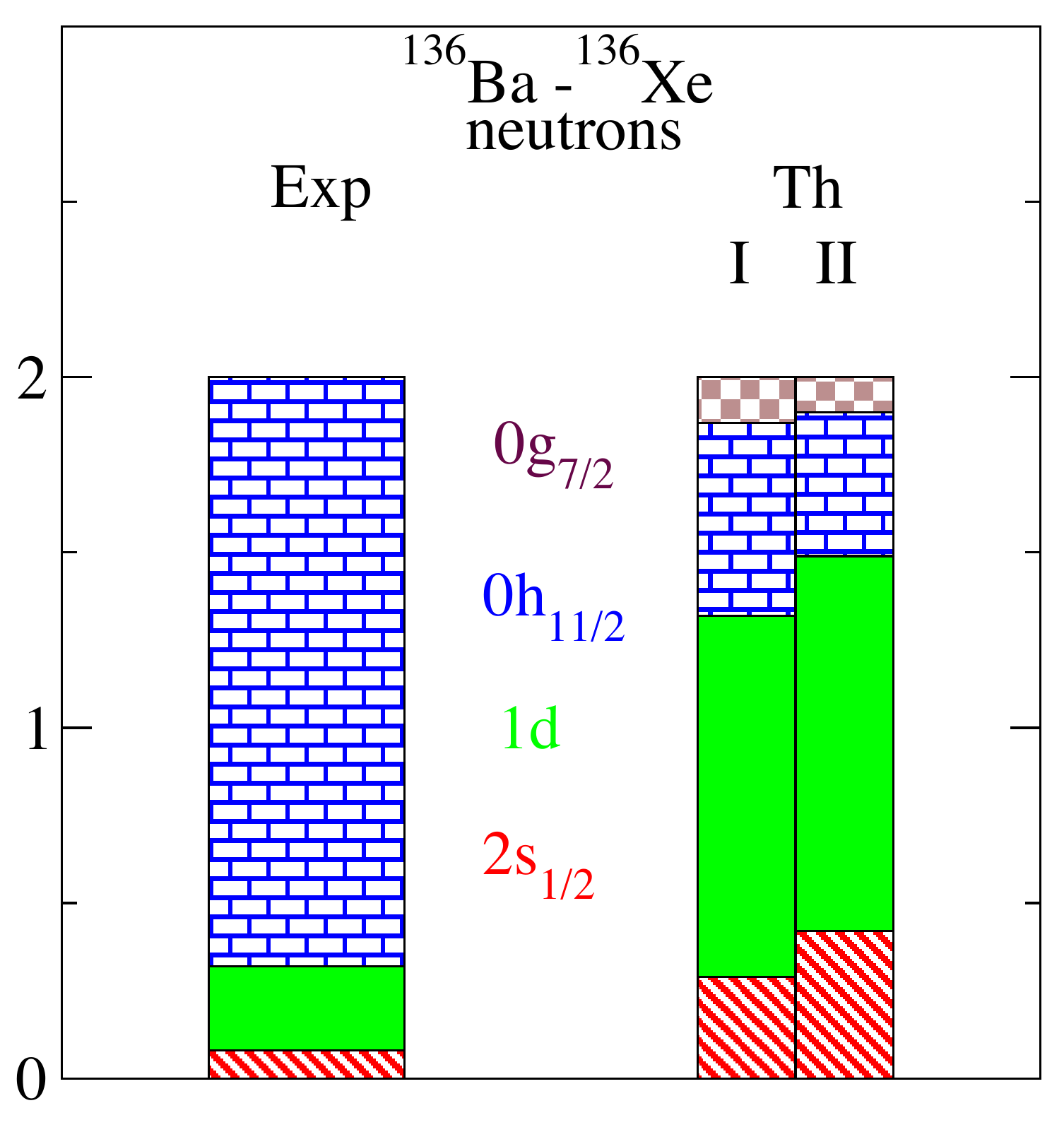}
\caption{Same as in Fig. \ref{130Teneut}, for the
  $^{136}{\rm Xe} \rightarrow ^{136}$Ba decay (see text for
  details).}
\label{136Xeneut}
\end{figure}
\end{center}

Finally, in Figs. \ref{136Xeprot} and \ref{136Xeneut} the theoretical
occupancies of valence nucleons in the ground states of the parent and
grand-daughter nuclei are shown and compared with those obtained in
Ref. \cite{Kay13,Entwisle16} from the experimental cross sections of
the $(d,^3{\rm He})$ and $(\alpha,^3{\rm  He})$ reactions.
The poor reproduction of the experimental neutron vacancies, as can be
seen in Fig. \ref{136Xeneut}, is due to the fact that in our model
space the neutron component of $^{136}$Xe is frozen, having its 32
valence neutrons totally filled the 50-82 shell.

\section{Summary and Outlook}

In the present work, we have presented the results of a realistic SM
calculation of GT decay properties for $^{130}$Te and $^{136}$Xe.
Our aim has been to test an approach to the calculation of the NME of
the $0\nu\beta\beta$ decay of these nuclei, where the SM Hamiltonian
and the related transition operators are derived, starting from a
realistic $NN$ potential, from the many-body theory.
This means that the need to resort to empirical parameters is
drastically reduced, so enhancing the predictiveness of the nuclear
structure calculations.

The first step toward this goal has been to test the reliability of
our theoretical framework, that cannot be done for the
$0\nu\beta\beta$. 
In particular we have calculated the GT strengths and the NMEs of the
$2\nu\beta\beta$, and compared the results with the available
experimental ones.
This is reported in Section \ref{results}, and the overall agreement
with the data is quite good.
As summarized in Table \ref{summaryGT}, the quality of our results is
similar to, and even better than, that obtained with recent
calculations available in literature, which employ SM parameters that
have been empirically fitted to reproduce some selected observables.

Our results are encouraging for our next steps towards an
(almost) parameter-free calculation of the $0\nu\beta\beta$ NME of
$^{130}$Te and $^{136}$Xe, making us confident of a positive outcome
of a fully microscopical approach to this problem.

Finally, it should be pointed out that in our calculations the
renormalization of the bare operators by way of the many-body
perturbation theory takes into account the degrees of freedom that
are not explicitly included within the reduced SM model space.

There are also two other main effects that should be included in the
renormalization of the GT operators, namely the blocking effect and
the role of the subnucleonic degrees of freedom.

\begin{table}[ht]
\caption{Experimental and calculated GT strengths and
  $2\nu\beta\beta$-decay NME (in MeV$^{-1}$) for $^{130}$Te and
  $^{136}$Xe.}
\begin{ruledtabular}
\begin{tabular}{ccccc}
\label{summaryGT}
 Nucleus & ~ & Expt. & I & II \\
\colrule
 ~ & ~ & ~& ~ & ~ \\
 $^{130}$Te   &          ~                     & ~ & ~ & ~ \\
~                 & GT strength & $0.746 \pm 0.045$ & 0.842 & 0.873  \\
~                 & NME & $0.034 \pm 0.003$ & 0.044 & 0.046  \\
 $^{136}$Xe   &          ~                     & ~ & ~ & ~ \\
~                 & GT strength & $1.33 \pm 0.07$ & 0.94 & 1.13 \\
~                 & NME & $0.0218 \pm 0.0003$ & 0.0285 & 0.0287  \\
\end{tabular}
\end{ruledtabular}
\end{table}

The blocking effect is responsible for taking into account the
correlations among the active nucleons in systems with many
interacting valence particles, within the derivation of the effective
operators.
We are currently investigating the role played by these correlations
in the calculation of GT and $2\nu\beta\beta$ properties.

Another contribution to the renormalization of the GT operators is
associated to the quark structure of nucleons. 
Since a realistic $NN$ potential is our starting point, we do not
consider in such a picture the role played by the nucleon resonances
($\Delta,N^{\ast},\cdots$) - that are also responsible for three-nucleon
forces - whose contribution should lead to renormalized values of
$g_A$ and $g_V$.

Nowadays, the derivation of nuclear potentials by way of the chiral
perturbation theory \cite{EGM05,ME11} allows a consistent treatment of
this approach to the renormalization of the axial- and vector-current
constants, that has been already explored in Ref. \cite{Menendez11}.
We are investigating this subject, which will be the topic of a
forthcoming paper.

\section*{Acknowledgments}

The authors gratefully acknowledge useful comments and suggestions
from Francesco Iachello and Frederic Nowacki.

\bibliographystyle{apsrev}
\bibliography{biblio.bib}

\newpage
\section*{Appendix}
\renewcommand{\thetable}{A.\Roman{table}}
\setcounter{table}{0}

\begin{table}[H]
\caption{Proton-proton, neutron-neutron, and proton-neutron matrix
  elements (in MeV) derived for calculations in model space
  $0g_{7/2},1d_{5/2},1d_{3/2},2s_{1/2},0h_{11/2}$. They are
  antisymmetrized, and normalized by a factor $1/ \sqrt{ (1 +
    \delta_{j_aj_b})(1 + \delta_{j_cj_d})}$.}
\begin{ruledtabular}
\begin{tabular}{cccc}
$n_a l_a j_a ~ n_b l_b j_b ~ n_c l_c j_c ~ n_d l_d j_d $ & $J$ & $T_z$
  &  TBME \\
\colrule
 $ 0g_{ 7/2}~ 0g_{ 7/2}~ 0g_{ 7/2}~ 0g_{ 7/2}$ &  0 &  1 & -0.426 \\
 $ 0g_{ 7/2}~ 0g_{ 7/2}~ 1d_{ 5/2}~ 1d_{ 5/2}$ &  0 &  1 & -0.660 \\
 $ 0g_{ 7/2}~ 0g_{ 7/2}~ 1d_{ 3/2}~ 1d_{ 3/2}$ &  0 &  1 & -0.333 \\
 $ 0g_{ 7/2}~ 0g_{ 7/2}~ 2s_{ 1/2}~ 2s_{ 1/2}$ &  0 &  1 & -0.253 \\
 $ 0g_{ 7/2}~ 0g_{ 7/2}~ 0h_{11/2}~ 0h_{11/2}$ &  0 &  1 &  1.751 \\
 $ 1d_{ 5/2}~ 1d_{ 5/2}~ 0g_{ 7/2}~ 0g_{ 7/2}$ &  0 &  1 & -0.660 \\
 $ 1d_{ 5/2}~ 1d_{ 5/2}~ 1d_{ 5/2}~ 1d_{ 5/2}$ &  0 &  1 & -0.394 \\
\end{tabular}
\end{ruledtabular}
\label{tbme}
\end{table}

\begin{table}[H]
\begin{ruledtabular}
\begin{tabular}{cccc}
\colrule
 $ 1d_{ 5/2}~ 1d_{ 5/2}~ 0g_{ 7/2}~ 0g_{ 7/2}$ &  2 &  1 & -0.105 \\
 $ 1d_{ 5/2}~ 1d_{ 5/2}~ 0g_{ 7/2}~ 1d_{ 5/2}$ &  2 &  1 &  0.034 \\
 $ 1d_{ 5/2}~ 1d_{ 5/2}~ 0g_{ 7/2}~ 1d_{ 3/2}$ &  2 &  1 & -0.114 \\
 $ 1d_{ 5/2}~ 1d_{ 5/2}~ 1d_{ 5/2}~ 1d_{ 5/2}$ &  2 &  1 & -0.026 \\
 $ 1d_{ 5/2}~ 1d_{ 5/2}~ 1d_{ 5/2}~ 1d_{ 3/2}$ &  2 &  1 & -0.179 \\
 $ 1d_{ 5/2}~ 1d_{ 5/2}~ 1d_{ 5/2}~ 2s_{ 1/2}$ &  2 &  1 & -0.246 \\
 $ 1d_{ 5/2}~ 1d_{ 5/2}~ 1d_{ 3/2}~ 1d_{ 3/2}$ &  2 &  1 & -0.243 \\
 $ 1d_{ 5/2}~ 1d_{ 5/2}~ 1d_{ 3/2}~ 2s_{ 1/2}$ &  2 &  1 &  0.200 \\
\end{tabular}
\end{ruledtabular}
\label{tbme}
\end{table}

\begin{table}[H]
\begin{ruledtabular}
\begin{tabular}{cccc}
\colrule
 $ 1d_{ 5/2}~ 2s_{ 1/2}~ 0g_{ 7/2}~ 2s_{ 1/2}$ &  3 &  1 &  0.024 \\
 $ 1d_{ 5/2}~ 2s_{ 1/2}~ 1d_{ 5/2}~ 2s_{ 1/2}$ &  3 &  1 &  0.250 \\
 $ 0g_{ 7/2}~ 0g_{ 7/2}~ 0g_{ 7/2}~ 0g_{ 7/2}$ &  4 &  1 &  0.219 \\
 $ 0g_{ 7/2}~ 0g_{ 7/2}~ 0g_{ 7/2}~ 1d_{ 5/2}$ &  4 &  1 &  0.049 \\
 $ 0g_{ 7/2}~ 0g_{ 7/2}~ 0g_{ 7/2}~ 1d_{ 3/2}$ &  4 &  1 & -0.068 \\
 $ 0g_{ 7/2}~ 0g_{ 7/2}~ 0g_{ 7/2}~ 2s_{ 1/2}$ &  4 &  1 &  0.031 \\
 $ 0g_{ 7/2}~ 0g_{ 7/2}~ 1d_{ 5/2}~ 1d_{ 5/2}$ &  4 &  1 & -0.058 \\
 $ 0g_{ 7/2}~ 0g_{ 7/2}~ 1d_{ 5/2}~ 1d_{ 3/2}$ &  4 &  1 & -0.146 \\
\end{tabular}
\end{ruledtabular}
\label{tbme}
\end{table}

\begin{table}[H]
\begin{ruledtabular}
\begin{tabular}{cccc}
\colrule
 $ 1d_{ 3/2}~ 0h_{11/2}~ 1d_{ 5/2}~ 0h_{11/2}$ &  4 &  1 &  0.000 \\
 $ 1d_{ 3/2}~ 0h_{11/2}~ 1d_{ 3/2}~ 0h_{11/2}$ &  4 &  1 &  0.271 \\
 $ 0g_{ 7/2}~ 0h_{11/2}~ 0g_{ 7/2}~ 0h_{11/2}$ &  5 &  1 &  0.187 \\
 $ 0g_{ 7/2}~ 0h_{11/2}~ 1d_{ 5/2}~ 0h_{11/2}$ &  5 &  1 &  0.088 \\
 $ 0g_{ 7/2}~ 0h_{11/2}~ 1d_{ 3/2}~ 0h_{11/2}$ &  5 &  1 & -0.052 \\
 $ 0g_{ 7/2}~ 0h_{11/2}~ 2s_{ 1/2}~ 0h_{11/2}$ &  5 &  1 &  0.046 \\
 $ 1d_{ 5/2}~ 0h_{11/2}~ 0g_{ 7/2}~ 0h_{11/2}$ &  5 &  1 &  0.088 \\
 $ 1d_{ 5/2}~ 0h_{11/2}~ 1d_{ 5/2}~ 0h_{11/2}$ &  5 &  1 &  0.111 \\
\end{tabular}
\end{ruledtabular}
\label{tbme}
\end{table}

\begin{table}[H]
\begin{ruledtabular}
\begin{tabular}{cccc}
\colrule
 $ 0h_{11/2}~ 0h_{11/2}~ 1d_{ 5/2}~ 1d_{ 5/2}$ &  0 & -1 &  0.717 \\
 $ 0h_{11/2}~ 0h_{11/2}~ 1d_{ 3/2}~ 1d_{ 3/2}$ &  0 & -1 &  0.713 \\
 $ 0h_{11/2}~ 0h_{11/2}~ 2s_{ 1/2}~ 2s_{ 1/2}$ &  0 & -1 &  0.399 \\
 $ 0h_{11/2}~ 0h_{11/2}~ 0h_{11/2}~ 0h_{11/2}$ &  0 & -1 & -0.915 \\
 $ 0g_{ 7/2}~ 1d_{ 5/2}~ 0g_{ 7/2}~ 1d_{ 5/2}$ &  1 & -1 &  0.024 \\
 $ 0g_{ 7/2}~ 1d_{ 5/2}~ 1d_{ 5/2}~ 1d_{ 3/2}$ &  1 & -1 & -0.033 \\
 $ 0g_{ 7/2}~ 1d_{ 5/2}~ 1d_{ 3/2}~ 2s_{ 1/2}$ &  1 & -1 & -0.017 \\
 $ 1d_{ 5/2}~ 1d_{ 3/2}~ 0g_{ 7/2}~ 1d_{ 5/2}$ &  1 & -1 & -0.033 \\
\end{tabular}
\end{ruledtabular}
\label{tbme}
\end{table}

\begin{table}[H]
\begin{ruledtabular}
\begin{tabular}{cccc}
\colrule
 $ 1d_{ 3/2}~ 1d_{ 3/2}~ 0h_{11/2}~ 0h_{11/2}$ &  2 & -1 &  0.134 \\
 $ 1d_{ 3/2}~ 2s_{ 1/2}~ 0g_{ 7/2}~ 0g_{ 7/2}$ &  2 & -1 &  0.093 \\
 $ 1d_{ 3/2}~ 2s_{ 1/2}~ 0g_{ 7/2}~ 1d_{ 5/2}$ &  2 & -1 & -0.024 \\
 $ 1d_{ 3/2}~ 2s_{ 1/2}~ 0g_{ 7/2}~ 1d_{ 3/2}$ &  2 & -1 &  0.186 \\
 $ 1d_{ 3/2}~ 2s_{ 1/2}~ 1d_{ 5/2}~ 1d_{ 5/2}$ &  2 & -1 &  0.218 \\
 $ 1d_{ 3/2}~ 2s_{ 1/2}~ 1d_{ 5/2}~ 1d_{ 3/2}$ &  2 & -1 &  0.238 \\
 $ 1d_{ 3/2}~ 2s_{ 1/2}~ 1d_{ 5/2}~ 2s_{ 1/2}$ &  2 & -1 &  0.460 \\
 $ 1d_{ 3/2}~ 2s_{ 1/2}~ 1d_{ 3/2}~ 2s_{ 1/2}$ &  2 & -1 & -0.159 \\
\end{tabular}
\end{ruledtabular}
\label{tbme}
\end{table}

\begin{table}[H]
\begin{ruledtabular}
\begin{tabular}{cccc}
\colrule
 $ 1d_{ 5/2}~ 1d_{ 3/2}~ 0g_{ 7/2}~ 1d_{ 5/2}$ &  4 & -1 &  0.127 \\
 $ 1d_{ 5/2}~ 1d_{ 3/2}~ 0g_{ 7/2}~ 1d_{ 3/2}$ &  4 & -1 & -0.101 \\
 $ 1d_{ 5/2}~ 1d_{ 3/2}~ 0g_{ 7/2}~ 2s_{ 1/2}$ &  4 & -1 &  0.149 \\
 $ 1d_{ 5/2}~ 1d_{ 3/2}~ 1d_{ 5/2}~ 1d_{ 3/2}$ &  4 & -1 & -0.506 \\
 $ 1d_{ 5/2}~ 1d_{ 3/2}~ 0h_{11/2}~ 0h_{11/2}$ &  4 & -1 &  0.208 \\
 $ 0h_{11/2}~ 0h_{11/2}~ 0g_{ 7/2}~ 0g_{ 7/2}$ &  4 & -1 &  0.203 \\
 $ 0h_{11/2}~ 0h_{11/2}~ 0g_{ 7/2}~ 1d_{ 5/2}$ &  4 & -1 & -0.164 \\
 $ 0h_{11/2}~ 0h_{11/2}~ 0g_{ 7/2}~ 1d_{ 3/2}$ &  4 & -1 &  0.161 \\
\end{tabular}
\end{ruledtabular}
\label{tbme}
\end{table}

\begin{table}[H]
\begin{ruledtabular}
\begin{tabular}{cccc}
\colrule
 $ 2s_{ 1/2}~ 0h_{11/2}~ 1d_{ 3/2}~ 0h_{11/2}$ &  6 & -1 & -0.002 \\
 $ 2s_{ 1/2}~ 0h_{11/2}~ 2s_{ 1/2}~ 0h_{11/2}$ &  6 & -1 &  0.101 \\
 $ 0g_{ 7/2}~ 0h_{11/2}~ 0g_{ 7/2}~ 0h_{11/2}$ &  7 & -1 & -0.178 \\
 $ 0g_{ 7/2}~ 0h_{11/2}~ 1d_{ 5/2}~ 0h_{11/2}$ &  7 & -1 &  0.117 \\
 $ 0g_{ 7/2}~ 0h_{11/2}~ 1d_{ 3/2}~ 0h_{11/2}$ &  7 & -1 & -0.176 \\
 $ 1d_{ 5/2}~ 0h_{11/2}~ 0g_{ 7/2}~ 0h_{11/2}$ &  7 & -1 &  0.117 \\
 $ 1d_{ 5/2}~ 0h_{11/2}~ 1d_{ 5/2}~ 0h_{11/2}$ &  7 & -1 & -0.038 \\
 $ 1d_{ 5/2}~ 0h_{11/2}~ 1d_{ 3/2}~ 0h_{11/2}$ &  7 & -1 &  0.244 \\
\end{tabular}
\end{ruledtabular}
\end{table}

\begin{table}[H]
\begin{ruledtabular}
\begin{tabular}{cccc}
\colrule
 $ 1d_{ 5/2}~ 0g_{ 7/2}~ 1d_{ 3/2}~ 2s_{ 1/2}$ &  1 &  0 &  0.222 \\
 $ 1d_{ 5/2}~ 0g_{ 7/2}~ 2s_{ 1/2}~ 1d_{ 3/2}$ &  1 &  0 & -0.230 \\
 $ 1d_{ 5/2}~ 0g_{ 7/2}~ 2s_{ 1/2}~ 2s_{ 1/2}$ &  1 &  0 &  0.079 \\
 $ 1d_{ 5/2}~ 0g_{ 7/2}~ 0h_{11/2}~ 0h_{11/2}$ &  1 &  0 &  0.168 \\
 $ 1d_{ 5/2}~ 1d_{ 5/2}~ 0g_{ 7/2}~ 0g_{ 7/2}$ &  1 &  0 &  0.134 \\
 $ 1d_{ 5/2}~ 1d_{ 5/2}~ 0g_{ 7/2}~ 1d_{ 5/2}$ &  1 &  0 & -0.013 \\
 $ 1d_{ 5/2}~ 1d_{ 5/2}~ 1d_{ 5/2}~ 1d_{ 5/2}$ &  1 &  0 & -0.313 \\
 $ 1d_{ 5/2}~ 1d_{ 5/2}~ 1d_{ 5/2}~ 1d_{ 3/2}$ &  1 &  0 &  0.741 \\
\end{tabular}
\end{ruledtabular}
\end{table}

\begin{table}[H]
\begin{ruledtabular}
\begin{tabular}{cccc}
\colrule
 $ 2s_{ 1/2}~ 2s_{ 1/2}~ 1d_{ 3/2}~ 1d_{ 5/2}$ &  1 &  0 & -0.465 \\
 $ 2s_{ 1/2}~ 2s_{ 1/2}~ 1d_{ 3/2}~ 1d_{ 3/2}$ &  1 &  0 & -0.003 \\
 $ 2s_{ 1/2}~ 2s_{ 1/2}~ 1d_{ 3/2}~ 2s_{ 1/2}$ &  1 &  0 & -0.125 \\
 $ 2s_{ 1/2}~ 2s_{ 1/2}~ 2s_{ 1/2}~ 2s_{ 1/2}$ &  1 &  0 & -1.209 \\
 $ 2s_{ 1/2}~ 2s_{ 1/2}~ 0h_{11/2}~ 0h_{11/2}$ &  1 &  0 &  0.222 \\
 $ 0h_{11/2}~ 0h_{11/2}~ 0g_{ 7/2}~ 0g_{ 7/2}$ &  1 &  0 & -0.799 \\
 $ 0h_{11/2}~ 0h_{11/2}~ 0g_{ 7/2}~ 1d_{ 5/2}$ &  1 &  0 & -0.196 \\
 $ 0h_{11/2}~ 0h_{11/2}~ 1d_{ 5/2}~ 0g_{ 7/2}$ &  1 &  0 &  0.168 \\
\end{tabular}
\end{ruledtabular}
\end{table}

\begin{table}[H]
\begin{ruledtabular}
\begin{tabular}{cccc}
\colrule
 $ 1d_{ 5/2}~ 1d_{ 5/2}~ 0g_{ 7/2}~ 0g_{ 7/2}$ &  2 &  0 & -0.107 \\
 $ 1d_{ 5/2}~ 1d_{ 5/2}~ 0g_{ 7/2}~ 1d_{ 5/2}$ &  2 &  0 &  0.022 \\
 $ 1d_{ 5/2}~ 1d_{ 5/2}~ 0g_{ 7/2}~ 1d_{ 3/2}$ &  2 &  0 & -0.042 \\
 $ 1d_{ 5/2}~ 1d_{ 5/2}~ 1d_{ 5/2}~ 1d_{ 5/2}$ &  2 &  0 & -0.290 \\
 $ 1d_{ 5/2}~ 1d_{ 5/2}~ 1d_{ 5/2}~ 1d_{ 3/2}$ &  2 &  0 & -0.119 \\
 $ 1d_{ 5/2}~ 1d_{ 5/2}~ 1d_{ 5/2}~ 2s_{ 1/2}$ &  2 &  0 & -0.180 \\
 $ 1d_{ 5/2}~ 1d_{ 5/2}~ 1d_{ 3/2}~ 0g_{ 7/2}$ &  2 &  0 & -0.062 \\
 $ 1d_{ 5/2}~ 1d_{ 5/2}~ 1d_{ 3/2}~ 1d_{ 5/2}$ &  2 &  0 &  0.111 \\
\end{tabular}
\end{ruledtabular}
\end{table}

\begin{table}[H]
\begin{ruledtabular}
\begin{tabular}{cccc}
\colrule
 $ 1d_{ 3/2}~ 1d_{ 3/2}~ 1d_{ 5/2}~ 2s_{ 1/2}$ &  2 &  0 & -0.146 \\
 $ 1d_{ 3/2}~ 1d_{ 3/2}~ 1d_{ 3/2}~ 1d_{ 3/2}$ &  2 &  0 & -0.047 \\
 $ 1d_{ 3/2}~ 1d_{ 3/2}~ 1d_{ 3/2}~ 2s_{ 1/2}$ &  2 &  0 &  0.065 \\
 $ 1d_{ 3/2}~ 1d_{ 3/2}~ 2s_{ 1/2}~ 1d_{ 5/2}$ &  2 &  0 & -0.132 \\
 $ 1d_{ 3/2}~ 1d_{ 3/2}~ 2s_{ 1/2}~ 1d_{ 3/2}$ &  2 &  0 & -0.068 \\
 $ 1d_{ 3/2}~ 1d_{ 3/2}~ 0h_{11/2}~ 0h_{11/2}$ &  2 &  0 &  0.134 \\
 $ 1d_{ 3/2}~ 2s_{ 1/2}~ 0g_{ 7/2}~ 0g_{ 7/2}$ &  2 &  0 &  0.060 \\
 $ 1d_{ 3/2}~ 2s_{ 1/2}~ 0g_{ 7/2}~ 1d_{ 5/2}$ &  2 &  0 & -0.078 \\
\end{tabular}
\end{ruledtabular}
\end{table}

\begin{table}[H]
\begin{ruledtabular}
\begin{tabular}{cccc}
\colrule
 $ 0g_{ 7/2}~ 0g_{ 7/2}~ 2s_{ 1/2}~ 0g_{ 7/2}$ &  3 &  0 &  0.077 \\
 $ 0g_{ 7/2}~ 0g_{ 7/2}~ 2s_{ 1/2}~ 1d_{ 5/2}$ &  3 &  0 & -0.005 \\
 $ 0g_{ 7/2}~ 0g_{ 7/2}~ 0h_{11/2}~ 0h_{11/2}$ &  3 &  0 & -0.344 \\
 $ 0g_{ 7/2}~ 1d_{ 5/2}~ 0g_{ 7/2}~ 1d_{ 5/2}$ &  3 &  0 & -0.184 \\
 $ 0g_{ 7/2}~ 1d_{ 5/2}~ 0g_{ 7/2}~ 1d_{ 3/2}$ &  3 &  0 & -0.136 \\
 $ 0g_{ 7/2}~ 1d_{ 5/2}~ 0g_{ 7/2}~ 2s_{ 1/2}$ &  3 &  0 & -0.212 \\
 $ 0g_{ 7/2}~ 1d_{ 5/2}~ 1d_{ 5/2}~ 0g_{ 7/2}$ &  3 &  0 &  0.236 \\
 $ 0g_{ 7/2}~ 1d_{ 5/2}~ 1d_{ 5/2}~ 1d_{ 5/2}$ &  3 &  0 & -0.093 \\
\end{tabular}
\end{ruledtabular}
\end{table}

\begin{table}[H]
\begin{ruledtabular}
\begin{tabular}{cccc}
\colrule
 $ 1d_{ 5/2}~ 1d_{ 3/2}~ 0g_{ 7/2}~ 2s_{ 1/2}$ &  3 &  0 &  0.034 \\
 $ 1d_{ 5/2}~ 1d_{ 3/2}~ 1d_{ 5/2}~ 1d_{ 3/2}$ &  3 &  0 & -0.176 \\
 $ 1d_{ 5/2}~ 1d_{ 3/2}~ 1d_{ 5/2}~ 2s_{ 1/2}$ &  3 &  0 &  0.165 \\
 $ 1d_{ 5/2}~ 1d_{ 3/2}~ 1d_{ 3/2}~ 0g_{ 7/2}$ &  3 &  0 &  0.045 \\
 $ 1d_{ 5/2}~ 1d_{ 3/2}~ 1d_{ 3/2}~ 1d_{ 5/2}$ &  3 &  0 &  0.180 \\
 $ 1d_{ 5/2}~ 1d_{ 3/2}~ 1d_{ 3/2}~ 1d_{ 3/2}$ &  3 &  0 &  0.353 \\
 $ 1d_{ 5/2}~ 1d_{ 3/2}~ 2s_{ 1/2}~ 0g_{ 7/2}$ &  3 &  0 & -0.034 \\
 $ 1d_{ 5/2}~ 1d_{ 3/2}~ 2s_{ 1/2}~ 1d_{ 5/2}$ &  3 &  0 &  0.207 \\
\end{tabular}
\end{ruledtabular}
\end{table}

\begin{table}[H]
\begin{ruledtabular}
\begin{tabular}{cccc}
\colrule
 $ 2s_{ 1/2}~ 0g_{ 7/2}~ 1d_{ 3/2}~ 1d_{ 3/2}$ &  3 &  0 &  0.201 \\
 $ 2s_{ 1/2}~ 0g_{ 7/2}~ 2s_{ 1/2}~ 0g_{ 7/2}$ &  3 &  0 & -0.271 \\
 $ 2s_{ 1/2}~ 0g_{ 7/2}~ 2s_{ 1/2}~ 1d_{ 5/2}$ &  3 &  0 &  0.052 \\
 $ 2s_{ 1/2}~ 0g_{ 7/2}~ 0h_{11/2}~ 0h_{11/2}$ &  3 &  0 &  0.034 \\
 $ 2s_{ 1/2}~ 1d_{ 5/2}~ 0g_{ 7/2}~ 0g_{ 7/2}$ &  3 &  0 & -0.005 \\
 $ 2s_{ 1/2}~ 1d_{ 5/2}~ 0g_{ 7/2}~ 1d_{ 5/2}$ &  3 &  0 & -0.095 \\
 $ 2s_{ 1/2}~ 1d_{ 5/2}~ 0g_{ 7/2}~ 1d_{ 3/2}$ &  3 &  0 &  0.041 \\
 $ 2s_{ 1/2}~ 1d_{ 5/2}~ 0g_{ 7/2}~ 2s_{ 1/2}$ &  3 &  0 & -0.088 \\
\end{tabular}
\end{ruledtabular}
\end{table}

\begin{table}[H]
\begin{ruledtabular}
\begin{tabular}{cccc}
\colrule
 $ 1d_{ 5/2}~ 0g_{ 7/2}~ 0g_{ 7/2}~ 0g_{ 7/2}$ &  4 &  0 & -0.040 \\
 $ 1d_{ 5/2}~ 0g_{ 7/2}~ 0g_{ 7/2}~ 1d_{ 5/2}$ &  4 &  0 & -0.302 \\
 $ 1d_{ 5/2}~ 0g_{ 7/2}~ 0g_{ 7/2}~ 1d_{ 3/2}$ &  4 &  0 & -0.063 \\
 $ 1d_{ 5/2}~ 0g_{ 7/2}~ 0g_{ 7/2}~ 2s_{ 1/2}$ &  4 &  0 & -0.127 \\
 $ 1d_{ 5/2}~ 0g_{ 7/2}~ 1d_{ 5/2}~ 0g_{ 7/2}$ &  4 &  0 & -0.229 \\
 $ 1d_{ 5/2}~ 0g_{ 7/2}~ 1d_{ 5/2}~ 1d_{ 5/2}$ &  4 &  0 & -0.020 \\
 $ 1d_{ 5/2}~ 0g_{ 7/2}~ 1d_{ 5/2}~ 1d_{ 3/2}$ &  4 &  0 & -0.084 \\
 $ 1d_{ 5/2}~ 0g_{ 7/2}~ 1d_{ 3/2}~ 0g_{ 7/2}$ &  4 &  0 & -0.095 \\
\end{tabular}
\end{ruledtabular}
\end{table}

\begin{table}[H]
\begin{ruledtabular}
\begin{tabular}{cccc}
\colrule
 $ 0h_{11/2}~ 0h_{11/2}~ 1d_{ 3/2}~ 0g_{ 7/2}$ &  4 &  0 &  0.110 \\
 $ 0h_{11/2}~ 0h_{11/2}~ 1d_{ 3/2}~ 1d_{ 5/2}$ &  4 &  0 & -0.158 \\
 $ 0h_{11/2}~ 0h_{11/2}~ 2s_{ 1/2}~ 0g_{ 7/2}$ &  4 &  0 &  0.136 \\
 $ 0h_{11/2}~ 0h_{11/2}~ 0h_{11/2}~ 0h_{11/2}$ &  4 &  0 & -0.217 \\
 $ 0g_{ 7/2}~ 0g_{ 7/2}~ 0g_{ 7/2}~ 0g_{ 7/2}$ &  5 &  0 & -0.264 \\
 $ 0g_{ 7/2}~ 0g_{ 7/2}~ 0g_{ 7/2}~ 1d_{ 5/2}$ &  5 &  0 & -0.134 \\
 $ 0g_{ 7/2}~ 0g_{ 7/2}~ 0g_{ 7/2}~ 1d_{ 3/2}$ &  5 &  0 & -0.259 \\
 $ 0g_{ 7/2}~ 0g_{ 7/2}~ 1d_{ 5/2}~ 0g_{ 7/2}$ &  5 &  0 &  0.124 \\
\end{tabular}
\end{ruledtabular}
\end{table}

\begin{table}[H]
\begin{ruledtabular}
\begin{tabular}{cccc}
\colrule
 $ 0h_{11/2}~ 0h_{11/2}~ 0h_{11/2}~ 0h_{11/2}$ &  9 &  0 & -0.350 \\
 $ 0h_{11/2}~ 0h_{11/2}~ 0h_{11/2}~ 0h_{11/2}$ & 10 &  0 &  0.028 \\
 $ 0h_{11/2}~ 0h_{11/2}~ 0h_{11/2}~ 0h_{11/2}$ & 11 &  0 & -1.148 \\
 $ 0g_{ 7/2}~ 0h_{11/2}~ 0g_{ 7/2}~ 0h_{11/2}$ &  2 &  0 & -1.420 \\
 $ 0g_{ 7/2}~ 0h_{11/2}~ 0h_{11/2}~ 0g_{ 7/2}$ &  2 &  0 &  1.066 \\
 $ 0h_{11/2}~ 0g_{ 7/2}~ 0g_{ 7/2}~ 0h_{11/2}$ &  2 &  0 &  1.066 \\
 $ 0h_{11/2}~ 0g_{ 7/2}~ 0h_{11/2}~ 0g_{ 7/2}$ &  2 &  0 & -1.349 \\
 $ 0g_{ 7/2}~ 0h_{11/2}~ 0g_{ 7/2}~ 0h_{11/2}$ &  3 &  0 & -0.746 \\
\end{tabular}
\end{ruledtabular}
\end{table}

\begin{table}[H]
\begin{ruledtabular}
\begin{tabular}{cccc}
\colrule
 $ 1d_{ 5/2}~ 0h_{11/2}~ 0h_{11/2}~ 1d_{ 3/2}$ &  5 &  0 & -0.047 \\
 $ 1d_{ 5/2}~ 0h_{11/2}~ 0h_{11/2}~ 2s_{ 1/2}$ &  5 &  0 & -0.055 \\
 $ 1d_{ 3/2}~ 0h_{11/2}~ 0g_{ 7/2}~ 0h_{11/2}$ &  5 &  0 & -0.240 \\
 $ 1d_{ 3/2}~ 0h_{11/2}~ 1d_{ 5/2}~ 0h_{11/2}$ &  5 &  0 &  0.107 \\
 $ 1d_{ 3/2}~ 0h_{11/2}~ 1d_{ 3/2}~ 0h_{11/2}$ &  5 &  0 & -0.212 \\
 $ 1d_{ 3/2}~ 0h_{11/2}~ 2s_{ 1/2}~ 0h_{11/2}$ &  5 &  0 &  0.222 \\
 $ 1d_{ 3/2}~ 0h_{11/2}~ 0h_{11/2}~ 0g_{ 7/2}$ &  5 &  0 & -0.161 \\
 $ 1d_{ 3/2}~ 0h_{11/2}~ 0h_{11/2}~ 1d_{ 5/2}$ &  5 &  0 &  0.043 \\
\end{tabular}
\end{ruledtabular}
\end{table}

\begin{table}[H]
\begin{ruledtabular}
\begin{tabular}{cccc}
\colrule
 $ 2s_{ 1/2}~ 0h_{11/2}~ 1d_{ 5/2}~ 0h_{11/2}$ &  6 &  0 & -0.287 \\
 $ 2s_{ 1/2}~ 0h_{11/2}~ 1d_{ 3/2}~ 0h_{11/2}$ &  6 &  0 & -0.195 \\
 $ 2s_{ 1/2}~ 0h_{11/2}~ 2s_{ 1/2}~ 0h_{11/2}$ &  6 &  0 & -0.276 \\
 $ 2s_{ 1/2}~ 0h_{11/2}~ 0h_{11/2}~ 0g_{ 7/2}$ &  6 &  0 &  0.118 \\
 $ 2s_{ 1/2}~ 0h_{11/2}~ 0h_{11/2}~ 1d_{ 5/2}$ &  6 &  0 & -0.255 \\
 $ 2s_{ 1/2}~ 0h_{11/2}~ 0h_{11/2}~ 1d_{ 3/2}$ &  6 &  0 &  0.212 \\
 $ 2s_{ 1/2}~ 0h_{11/2}~ 0h_{11/2}~ 2s_{ 1/2}$ &  6 &  0 & -0.370 \\
 $ 0h_{11/2}~ 0g_{ 7/2}~ 0g_{ 7/2}~ 0h_{11/2}$ &  6 &  0 &  0.173 \\
\end{tabular}
\end{ruledtabular}
\end{table}

\begin{table}[H]
\begin{ruledtabular}
\begin{tabular}{cccc}
\colrule
 $ 0g_{ 7/2}~ 0h_{11/2}~ 0h_{11/2}~ 1d_{ 5/2}$ &  8 &  0 & -0.180 \\
 $ 1d_{ 5/2}~ 0h_{11/2}~ 0g_{ 7/2}~ 0h_{11/2}$ &  8 &  0 & -0.125 \\
 $ 1d_{ 5/2}~ 0h_{11/2}~ 1d_{ 5/2}~ 0h_{11/2}$ &  8 &  0 & -0.546 \\
 $ 1d_{ 5/2}~ 0h_{11/2}~ 0h_{11/2}~ 0g_{ 7/2}$ &  8 &  0 &  0.198 \\
 $ 1d_{ 5/2}~ 0h_{11/2}~ 0h_{11/2}~ 1d_{ 5/2}$ &  8 &  0 & -0.704 \\
 $ 0h_{11/2}~ 0g_{ 7/2}~ 0g_{ 7/2}~ 0h_{11/2}$ &  8 &  0 &  0.069 \\
 $ 0h_{11/2}~ 0g_{ 7/2}~ 1d_{ 5/2}~ 0h_{11/2}$ &  8 &  0 &  0.198 \\
 $ 0h_{11/2}~ 0g_{ 7/2}~ 0h_{11/2}~ 0g_{ 7/2}$ &  8 &  0 & -0.022 \\
 $ 0h_{11/2}~ 0g_{ 7/2}~ 0h_{11/2}~ 1d_{ 5/2}$ &  8 &  0 &  0.125 \\
 $ 0h_{11/2}~ 1d_{ 5/2}~ 0g_{ 7/2}~ 0h_{11/2}$ &  8 &  0 & -0.180 \\
 $ 0h_{11/2}~ 1d_{ 5/2}~ 1d_{ 5/2}~ 0h_{11/2}$ &  8 &  0 & -0.704 \\
 $ 0h_{11/2}~ 1d_{ 5/2}~ 0h_{11/2}~ 1d_{ 5/2}$ &  8 &  0 & -0.516 \\
 $ 0g_{ 7/2}~ 0h_{11/2}~ 0g_{ 7/2}~ 0h_{11/2}$ &  9 &  0 & -1.057 \\
 $ 0g_{ 7/2}~ 0h_{11/2}~ 0h_{11/2}~ 0g_{ 7/2}$ &  9 &  0 & -0.255 \\
 $ 0h_{11/2}~ 0g_{ 7/2}~ 0g_{ 7/2}~ 0h_{11/2}$ &  9 &  0 & -0.255 \\
 $ 0h_{11/2}~ 0g_{ 7/2}~ 0h_{11/2}~ 0g_{ 7/2}$ &  9 &  0 & -0.989 \\
\end{tabular}
\end{ruledtabular}
\end{table}

\end{document}